\documentclass[a4paper,12pt]{article}
\usepackage[centertags]{amsmath}
\usepackage{amsfonts}
\usepackage[]{epsfig}
\newcommand{\be}{\begin{equation}}
\newcommand{\ee}{\end{equation}}
\newcommand{\bea}{\begin{eqnarray}}
\newcommand{\eea}{\end{eqnarray}}

\newcommand{\bmat}{\left(\begin{array}}
\newcommand{\emat}{\end{array}\right)}

\title{}
\author{}
\begin{document}




\numberwithin{equation}{section}

\newcommand{\df}{\stackrel{\rm def}{=}}
\newcommand{\co}{{\scriptstyle \circ}}
\newcommand{\lb}{\lbrack}
\newcommand{\rb}{\rbrack}
\newcommand{\rn}[1]{\romannumeral #1}
\newcommand{\msc}[1]{\mbox{\scriptsize #1}}
\newcommand{\dsp}{\displaystyle}
\newcommand{\scs}[1]{{\scriptstyle #1}}

\newcommand{\ket}[1]{| #1 \rangle}
\newcommand{\bra}[1]{| #1 \langle}
\newcommand{\vac}{| \mbox{vac} \rangle }

\newcommand{\e}{\mbox{{\bf e}}}
\newcommand{\va}{\mbox{{\bf a}}}
\newcommand{\bc}{\mbox{{\bf C}}}

\newcommand{\com}{C\!\!\!\!|}

\newcommand{\br}{\mbox{{\bf R}}}
\newcommand{\bz}{\mbox{{\bf Z}}}
\newcommand{\bq}{\mbox{{\bf Q}}}
\newcommand{\bn}{\mbox{{\bf N}}}
\newcommand {\eqn}[1]{(\ref{#1})}

\newcommand{\cp}{\mbox{{\bf P}}^1}
\newcommand{\n}{\mbox{{\bf n}}}
\newcommand{\sbz}{\msc{{\bf Z}}}
\newcommand{\sn}{\msc{{\bf n}}}

\newcommand{\cleqn}{\setcounter{equation}{0}}



\def\theequation{\thesection.\arabic{equation}}
\makeatother

\def\oa{\bigcirc\!\!\!\! a}
\def\ob{\bigcirc\!\!\!\! b}
\def\oc{\bigcirc\!\!\!\! c}
\def\oi{\bigcirc\!\!\!\! i}
\def\oj{\bigcirc\!\!\!\! j}
\def\ok{\bigcirc\!\!\!\! k}
\def\ve{\vec e}\def\vk{\vec k}\def\vn{\vec n}\def\vp{\vec p}
\def\vr{\vec r}\def\vs{\vec s}\def\vt{\vec t}\def\vu{\vec u}
\def\vv{\vec v}\def\vx{\vec x}\def\vy{\vec y}\def\vz{\vec z}

\title{On modular properties of the  AdS$_3$ CFT}

\author{Walter H. Baron  $^{1,}$ \footnote{e-mail: w\_baron@iafe.uba.ar} \, and
  Carmen~A.~N\'u\~nez $^{1,~2,}$
\footnote{e-mail: carmen@iafe.uba.ar}$^\dag$}

\date{\small $^1$ Instituto de Astronom\'{\i}a y F\'{\i}sica del Espacio
  (CONICET-UBA).\\
C.~C.~67 - Suc.~28, 1428 Buenos Aires, Argentina \\ and \\
  $^2$ Departamento de F\'\i sica,
FCEN, Universidad de Buenos Aires. \\ Ciudad Universitaria, Pab. I,
 1428 Buenos Aires,
Argentina.}

\maketitle
\begin{abstract}
We study  modular properties of the AdS$_3$ WZNW model. Although the
Euclidean partition function is modular invariant,
the characters  on the Euclidean torus are ill-defined
and
their modular transformations 
are unknown.
We reconsider the  characters 
defined on the Lorentzian torus, 
focusing on their structure as distributions.
We find a generalized $S$ matrix,
depending on the sign of the real modular parameters,
which
 has two diagonal blocks
 and one off-diagonal block,
mixing discrete and continuous representations, that we fully determine. 
We then  explore the relations among the modular transformations, 
 the fusion algebra and the
boundary states. We
 explicitly construct Ishibashi states for the maximally symmetric
D-branes and
 show that  the  generalized $S$ matrix defines the
one-point functions associated to point-like 
and H$_2$ branes  as well as the
fusion rules of  the degenerate representations of 
SL(2,$\mathbb R$) appearing in the open string spectrum of the point-like 
D-branes,
through a generalized Verlinde theorem.

\end{abstract}
\newpage

\tableofcontents

\newpage
\section{Introduction}

The formulation of a consistent  string theory on AdS$_3$ is an active area
of research
since more than two decades ago.
 Besides allowing to 
understand important aspects of  strings propagating
on non trivial  backgrounds (see \cite{mo1}-\cite{mo3} and references therein),
 this theory offers a controlled setting where it is
possible to 
 verify  the AdS/CFT 
correspondence beyond the supergravity approximation 
 as well as to grasp several features
 of a non rational conformal field theory (RCFT) with
Lie algebra symmetry.

The worldsheet
theory describing strings on Lorentzian AdS$_3$
 is a WZNW model on the universal cover of the 
SL(2,$\mathbb R$) group manifold. The spectrum proposed
in \cite{mo1} was verified in \cite{mo2} through  the computation of the 
one-loop partition function on a Euclidean AdS$_3$ background 
at finite temperature.
Some correlation functions
were determined in \cite{mo3} and 
 the fusion rules establishing the
closure of the Hilbert space and
the unitarity of the full interacting string
theory were obtained in our previous work \cite{wc}.
We showed that the spectral flow symmetry of the model requires 
a truncation of the operator algebra 
whose physical origin has not been elucidated yet.
Although 
they  satisfy several essential
properties, the full consistency of the fusion
rules should follow from a proof of factorization and crossing symmetry of the
four-point functions, still unavailable.
The correlators
that have been analyzed
in the literature so far 
are based on the analytic continuation from those of the
better understood Euclidean version of the theory, the H$_3^+\equiv
\frac{{\rm SL}(2, {\mathbb C})}{{\rm SU}(2)}$ WZNW model \cite{tesch1, tesch3}.
But there are many subtleties in the relation 
between the Euclidean and Lorentzian models \cite{rib} and
 further work is necessary to put the fusion rules
 on a firmer
 ground.

In RCFT, a practical derivation of the fusion rules
 can be performed through the Verlinde
theorem \cite{verlinde}, often formulated as the statement that the $S$ matrix
of modular transformations diagonalizes the fusion rules. 
Moreover, besides leading to
 a Verlinde formula, 
the $S$ matrix 
allows a classification
of modular invariants and
a systematic study
of boundary states. 
It would be interesting to explore whether 
analogues of these properties
can be found
in the AdS$_3$ WZNW  model. 
However, the
 relations among fusion algebra, boundary states and
modular transformations are difficult to identify  and 
 have not been very convenient
in non compact models \cite{TJ}. 
In general, 
the characters  have an
intricate behaviour under the modular group \cite{stt}-\cite{t} and,
as is often the case in 
theories with discrete and continuous representations, these mix under 
$S$ transformations. 

In this paper we study the modular properties of
 the AdS$_3$ model.
Since the
characters of the relevant representations 
diverge and lack good modular properties, 
extended characters
have been introduced in
\cite{hhrs, HS}. 
Instead, here we reconsider 
the standard characters.
Due to the divergences, these were computed on the Lorentzian torus
 in \cite{mo1} and
it was shown that the 
partition function of the  H$_3^+$ model obtained in \cite{gawe}
is recovered after performing analytic continuation 
 and discarding contact terms.
In section 2,
we review (and redefine) these characters, focusing
 on their structure as distributions. We also consider
the characters of  degenerate representations 
of SL(2,$\mathbb R$), because they
 appear in the boundary
spectrum of point-like D-brane solutions. 

Then we study their modular transformations. We
find generalized modular maps
which play an important role in the microscopic
description of the  theory. Real modular 
parameters are crucial 
 to obtain an $S$ matrix which,
unlike those of the Euclidean models, 
 depends on the sign of the modulus. 
In section 3, we completely determine this
generalized $S$
 matrix, which has two diagonal blocks and one off-diagonal block
 mixing the characters of discrete 
and continuous 
representations.

In order to explore the properties of this modular matrix, in
section 4 we consider 
the  maximally symmetric D-branes of the model.
We explicitly construct the Ishibashi states 
and show that 
the coefficients of the boundary states turn out to be
determined from the generalized $S$ matrix, 
suggesting that a Verlinde-like formula
 could give some information on
 the spectrum of  open strings attached to certain D-branes.
Furthermore,  
we show in appendix C that a generalized Verlinde
formula 
reproduces the fusion rules of the finite dimensional
 degenerate representations of 
SL(2,$\mathbb R$) appearing in the boundary spectrum of the point-like
 D-branes.

Conclusions are offered
in section 5, where we compare our results with previous ones in the literature
and we also
 draw some directions for future work. 

For the benefit of the reader,
we include four appendices. In appendix
A we discuss the properties of the moduli space of the
Lorentzian torus. Some details of the calculations
leading to the generalized $S$ matrix are presented in appendix B.
A generalized Verlinde formula giving
 the fusion rules of the degenerate representations
is worked out  
in appendix C.
Finally, in appendix D we 
review the results 
 of the
one-point functions for maximally symmetric D-branes obtained in
\cite{israel} 
and translate them to our conventions, in order to compare
with the expressions obtained in the main body of the text.

\section{Characters on the Lorentzian torus}

The partition function of the AdS$_3$ WZNW model was computed 
on the Lorentzian torus in  \cite{mo1} because it diverges
on the Euclidean signature torus,
and it was shown that  a modular invariant expression is obtained
after analytic continuation of the modular parameters\footnote{The same 
expression was
independently  obtained  
 in \cite{kounnas} where the Euclidean version of AdS$_3$ was 
constructed 
from the axial coset
 SL(2,${\mathbb R}$)/U(1$)_A$, using path integral techniques.}. 
In this section we 
rederive the characters of the relevant representations and
stress some important issues 
related to the regions of convergence of the expressions involved,
focussing on their structure as distributions.

The characters 
are defined on the Lorentzian signature torus from the standard
expressions as
\bea
\chi_{{\cal V}_L}(\theta_-,\tau_-,u_-)={\rm Tr}_{{\cal V}_L} e^{2\pi 
i\tau_-(L_0-\frac{c}{24})}
e^{2\pi i\theta_- J_0^3}e^{\pi iu_- K}\, ,\nonumber\\
\chi_{{\cal V}_R}(\theta_+,\tau_+,u_+)={\rm Tr}_{{\cal V}_R} e^{2\pi 
i\tau_+(\bar L_0-\frac{\bar c}{24})}
e^{2\pi i\theta_+ \bar J_0^3}e^{\pi iu_+ K}\, ,\label{char}
\eea
where 
 $\tau_\pm, \theta_\pm, u_\pm$  are independent real parameters,
  $c=\bar c$ are the right- and left-moving central charges
and
$K$ is the central element of the affine algebra. 
The Euclidean version is obtained replacing
the real parameters by complex ones.
For completeness, a
 description of the moduli space of the Lorentzian torus is presented in 
appendix A.

The traces in (\ref{char}) are taken over the left and
right representation
modules of the
Hilbert space of the theory, ${\cal V}_L$ 
and  ${\cal V}_R$, repectively. 
The spectrum of the AdS$_3$ WZNW model determined in \cite{mo1}
 decomposes into direct products of the normalizable continuous and lowest
weight discrete representations of the universal cover of the
affine SL(2,${\mathbb R})$ algebra with level $k\in {\mathbb R}_{>2}$.
The lowest principal discrete representations $\hat{\mathcal D}_j^+
 \times \hat{\mathcal D}_j^+$
 contain the states $|j,m,\bar m>$ with $-\frac{k-1}{2}<j<-\frac12$,
 $m ,\bar m\in -j+{\mathbb Z}_{\ge 0}$ and their affine
descendants. The
 principal continuous 
representations $\hat{\mathcal C}_j^\alpha
\times \hat{\mathcal C}_j^\alpha$ contain the states  $|j,\alpha, m, \bar m>$ 
with 
$j\in -\frac12+i{\mathbb R}^+$, $\alpha\in[0,1)$, $m,\bar m\in
\alpha+{\mathbb Z}$,
and their affine descendants. 
The spectrum also includes the spectral flow images of these representations, 
which can be constructed with the spectral flow operators
$U_w, \bar U_{\bar w}$, defined by their action on the SL(2,$\mathbb R)$ currents $J^3, J^\pm$ as
\bea
\left\{\begin{array}{lcr}
U_{-w} J^3(z) U_{w}&=&J^3(z)+\frac k2 \frac wz~,\cr
&&\cr
U_{-w} J^\pm(z) U_{w}&=&z^{\mp w}J^\pm(z)~,
\end{array}\right.~~~~~~  
\left\{\begin{array}{lcr}
\bar U_{-\bar w} \bar J^3(\bar z) \bar U_{\bar w}&=&\bar J^3(\bar z)+
\frac k2 \frac{\bar w}{\bar z}~,\cr
&&\cr
\bar U_{-\bar w} \bar J^\pm(\bar z) \bar U_{\bar w}&=&\bar z^{\mp \bar w}\bar 
J^\pm(\bar z)~,
\end{array}\right.~~~~~~\label{wtrans}
\eea 
where $U_{-w}=U_w^{-1},~\bar U_{-\bar w}=\bar U_{\bar w}^{-1}$
and $w= \bar w \in {\mathbb Z}$.\footnote{
The right and left spectral flow numbers
 $w,
\bar w$ are  not necessarily equal
in the single cover of SL(2,${\mathbb R})$
 where $\bar w-w$ is the winding number 
around the compact closed timelike direction.} Using the Sugawara 
construction, the action of $U_w, \bar U_{\bar w}$ on the zero modes of the
Virasoro generators
 is found to be
\bea
U_{-w} L_0 U_{w}=L_0-wJ_0^3-\frac k4 w^2~,\qquad
\bar U_{-\bar w} \bar L_0 \bar U_{\bar w}=\bar L_0-\bar w\bar J_0^3-
\frac k4 \bar w^2~, \label{wl0}
\eea
and
 the eigenvalues of $L_0,\bar L_0$ are, in general, not bounded 
from below. 
For states in the discrete series it is often convenient to work with 
spectral flow images of both lowest and highest weight representations, which
are related by the identification $\hat{\cal D}_j^{+,w}
\equiv \hat{\cal D}_{-\frac k2-j}^{-,w+1}$.

In the remaining of
this section we review (and redefine) the complete set of 
characters of the relevant 
representations making up the spectrum of the
bulk AdS$_3$ conformal field theory and of the finite dimensional
 representations 
appearing in the open string spectrum of some 
brane solutions.  

To lighten notation, from now on  $\tau , \theta, u$ 
will denote  the 
real parameters 
$\tau_-, \theta_-, u_-$ and 
 the following compact notation will be used: 
$\chi_j^{\pm,w}:=\chi_{\hat{\mathcal D}_j^{\pm ,w}}$,
$\chi_j^{\alpha,w}:=
\chi_{\hat{\mathcal C}_j^{\alpha,w}}$.

\subsection{Discrete representations}

The naive computation of the characters (\ref{char}) for the
 discrete representations
leads to $\theta$ and $\tau$ dependent 
divergences. This is not a problem because the characters
 are typically not functions but distributions. Indeed,
similarly as the characters of the continuous representations, which contain
 a series of delta functions \cite{mo1},
those of the discrete representations need also be interpreted as 
distributions. 

Let us consider the distributions constructed from
 the series defining the characters of the discrete 
representations. Shifting  $\tau\rightarrow\tau+i\xi_1$ and
 $\theta\rightarrow\theta+i\xi_2^w$ in (\ref{char}), 
where $\xi_1,\xi_2^w$ are two real non vanishing parameters,
 a regular distribution can be defined.
Indeed,
the deformed characters of  discrete representations
in an arbitrary spectral flow
sector $w$ can be written in terms 
of those of unflowed representations as
\bea
&&\chi_{j,\xi_2^w,\xi_1}^{+,w}
(\theta, \tau,u)=
e^{i\pi k u}\sum_n
\epsilon_n <n|U_{-w}e^{2\pi i 
(\tau+i\xi_1)(L_0-\frac{c}{24})} e^{2\pi i(\theta+i\xi_2^w) J_0^3} U_{w}|n> \,
,
\nonumber
\eea
where $|n>$ is a complete orthonormal basis in $\hat{\mathcal D}_j^{+,0}$,
 with norm 
$\epsilon_n=\pm1$ (recall that
 this model is not unitary). 
Since $U_w$ is unitary, 
$U_w|n>$ defines an orthonormal basis in 
$\hat{\mathcal D}_j^{+,w}$ and from (\ref{wtrans}) one can rewrite 
\bea
\chi_{j,\xi_2^w,\xi_1}^{+,w}=
e^{i\pi k u} e^{-2\pi i\tau\frac k4w^2}e^{2\pi i\theta \frac k2 w}
\sum_n\epsilon_n<n|e^{2\pi i 
(\tau+i\xi_1)(L_0-\frac{c}{24})} e^{2\pi i(\theta-w\tau+ 
i(\xi_2^w-w\xi_1))J_0^3}|n>.~
\label{charfromw0}
\eea
Choosing an orthonormal basis of eigenvectors of $L_0$ and $J_0^3$, the
following behavior of the sum is easy to see
\bea
\chi_{j,\xi_2^w,\xi_1}^{+,w}\sim
 \sum_{N,n=0}^\infty \rho(n,N) ~e^{2\pi i [(1+w)\tau-
\theta +i\left((1+w)\xi_1-\xi_2^w\right)]N} 
e^{2\pi i[\theta-w\tau+i(\xi_2^w-w\xi_1)]n}\, ,\nonumber
\eea
where $\rho(n,N)$ gives the degeneracy of states. 
This expression is convergent for 
parameters in the ranges
\bea
\left\{\begin{array}{lcr}
~~~~~~~~~~~\xi_1>0\,,\cr
(1+w)\xi_1>\xi_2^w>w\xi_1 \, ,
\end{array}\right.\label{epsiloncond}
\eea
 and it gives 
\bea
\chi_{j,\xi_2^w,\xi_1}^{+,w}
 =
e^{i\pi k u} e^{-2\pi i(\tau+i\xi_1)\frac k4w^2}e^{2\pi i(\theta+i\xi_2^w) 
\frac k2 w} 
\frac{e^{-\frac{2\pi i (\tau+i\xi_1)}{k-2}(j+\frac12)^2}
e^{-2\pi i(\theta+i\xi_2^w-
w(\tau+i\xi_1))(j+\frac12)}}
{i\vartheta_{11}(\theta+i\xi_2^w-w(\tau+i\xi_1),\tau+i\xi_1)}\, .~~ ~~
\eea

This character defines a regular distribution and,
given that the series 
of regular 
distributions are continuous with respect to the weak limit, 
this implies 
\bea
\chi_j^{+,w}(\theta,\tau,u)&=&
e^{i\pi k u}~\frac{e^{-\frac{2\pi i \tau}{k-2}(j+\frac12-w\frac{k-2}{2})^2} 
e^{-2\pi i\theta(j+\frac12-w\frac{k-2}{2})}}{i\vartheta_{11} 
(\theta+i\epsilon_2^w,\tau+i\epsilon_1)}\, ,\label{chidiscw}
\eea
where we have used the identity
\bea
\vartheta_{11}\left(\theta+i\epsilon_2^w -w(\tau +i\epsilon_1),
\tau+i\epsilon_1\right)=(-)^w e^{-\pi i \tau w^2+2\pi i \theta w}
\vartheta_{11}(\theta +i\epsilon_2^w, \tau+i\epsilon_1)
\eea
and the $i\epsilon$'s denote the usual $i0$ prescriptions,
 constrained as the corresponding finite parameters in
(\ref{epsiloncond}),
which dictate how to avoid the
poles of $\vartheta_{11}^{-1}$ 
at $n\tau\in {\mathbb Z},~m\tau+\theta\in{\mathbb Z}$, 
for $n\in{\mathbb N},m\in{\mathbb Z}$.
These poles are easily seen in the following alternative expression for the 
elliptic theta function
\bea
\frac{1}{\vartheta_{11}(\theta+i\epsilon_2^w,\tau+i\epsilon_1)} &=&
\frac{-e^{-i\frac\pi4 \tau}}{\sin\left[\pi\left(\theta+i\epsilon_2^w\right)
\right]}\frac 1{
\prod_{n=1}^{\infty}\left[1-e^{2\pi in(\tau+i\epsilon_1)
}\right] } 
\nonumber\\
&& \times ~\frac 1{\prod_{n=1}^\infty \left
[1-e^{2\pi i(n\tau-\theta+i\epsilon_3^{n,w}})
\right]
\left[1-e^{2\pi i(n\tau+\theta+i\epsilon_4^{n,w})}\right]}\, ,\label{theta11b}
\eea
\bea 
{\rm with} 
\ \ \ \ \ \ \ 
\left\{\begin{array}{lcr}
\epsilon_3^{n,w}=n\epsilon_1-\epsilon_2^w\cr
\epsilon_4^{n,w}=n\epsilon_1+\epsilon_2^w
\end{array}\right.\, ,\label{epsilon34}
\eea
$i.e.$, $\epsilon_3^{n,w}>0~(<0)$ for $n\geq1+w ~(n\leq w)$ 
and $\epsilon_4^{n,w}>0~(<0)$ for $n\geq-w~ (n\leq -1-w)$.

Notice that, in the weak limit, one can take
 $\epsilon_1,\,\epsilon_2^w=0$ in the arguments of the exponential 
terms in (\ref{chidiscw}) because they are perfectly regular.

It is useful to rewrite (\ref{chidiscw}) using the identity 
(\ref{identdeltas}), which allows  to change the signs of  $\epsilon_2^w,\,
\epsilon_3^{n,w}$ and $\epsilon_4^{n,w}$,
in order to get the following expressions in terms of
only one parameter, say $\epsilon_2^{w'}$, with arbitrary $w'$:
\bea
\chi_j^{+,w < w'}(\theta,\tau,u)&=&(-)^{w} e^{i\pi ku}\frac{
e^{-\frac{2\pi i 
\tau}{k-2}(j+\frac12-w\frac{k-2}{2})^2}e^{-2\pi i\theta(j+\frac12-w\frac{k-2}
{2})}}{i\vartheta_{11}(\theta+i\epsilon_2^{w'},\tau+i\epsilon_1)}\cr\cr
&&-~(-)^{w} e^{i\pi ku}\frac{
e^{-\frac{2\pi i \tau}{k-2}(j+\frac12-w\frac{k-2}{2})^2} 
e^{-2\pi i\theta(j+\frac12-w\frac{k-2}{2})}}{\eta^3(\tau+i\epsilon_1)}\cr
&&\times ~\sum_{n=1+w}^{w'} 
(-)^ne^{2i\pi \tau \frac{n^2}{2}}\sum_{m=-\infty}^{\infty}(-)^m
\delta(\theta-n\tau+m)\, ,\label{chiw<w'}
\eea  
\bea
\chi_j^{+,w > w'}(\theta,\tau,0)&=&(-)^{w} e^{i\pi k u}
\frac{e^{-\frac{2\pi i \tau}{k-2}(j+\frac12-w\frac{k-2}{2})^2}
e^{-2\pi i\theta(j+\frac12-w\frac{k-2}{2})}}{i\vartheta_{11}
(\theta+i\epsilon_2^{w'},\tau+i\epsilon_1)}\cr\cr
&& +~(-)^{w} e^{i\pi ku}
\frac{e^{-\frac{2\pi i \tau}{k-2}(j+\frac12-w\frac{k-2}{2})^2}
e^{-2\pi i\theta(j+\frac12-w\frac{k-2}{2})}}{\eta^3(\tau+i\epsilon_1)} \cr
&&\times ~ \sum_{n=1+w'}^{w} 
(-)^n e^{2i\pi \tau \frac{n^2}{2}}\sum_{m=-\infty}^{\infty}(-)^m
\delta(\theta-n\tau+m)\, .\label{chiw>w'}
\eea

These 
expressions are in perfect agreement with the spectral flow symmetry, 
which implies $\chi_{j}^{+,w}(-\theta,\tau,u)=\,\chi_{-\frac k2-j}^{+,-w-1}
(\theta,\tau,u)$.   
They lead to the following contribution to the partition function  
\bea
Z_{\cal D}^{AdS_3}= \sqrt{
\frac{k-2}{2i(\tau_- -\tau_+)}}
\frac{e^{i\pi k(u_--u_+)}
e^{2\pi i\frac{k-2}{4}\frac{(\theta_--\theta_+)^2}{\tau_- -\tau_+}}
}{
\vartheta_{11}(\theta_-+i\epsilon^0_2,\tau_-+i\epsilon_1)
\vartheta_{11}^*(\theta_+-i\epsilon^0_2,\tau_+-i\epsilon_1)}
~+~\dots\, , \label{parf}
\eea
where the
ellipses stand for the contributions of the contact terms. 
This expression differs formally from the equivalent one in \cite{mo1}, 
where no $\epsilon$ prescription or contact terms were considered. 
Nevertheless, 
the ultimate goal in \cite{mo1}
was to 
 reproduce the Euclidean partition function
continuing the modular parameters 
away from the real axes and discarding
contact terms such as those of the 
characters of the continuous representations.

\subsection{Continuous representations}

A similar analysis can be performed for the characters of the
continuous representations. 
Using (\ref{charfromw0}), one can compute these characters in terms of 
those of the unflowed continuous representations. The result is  
\bea
\chi_{j}^{\alpha,w}&=&e^{i\pi k u}
\frac{-2\sin[\pi (\theta-w\tau)]e^{-2\pi i\tau\frac k4w^2}
e^{2\pi i\theta \frac k2 w} 
e^{-\frac{2\pi i \tau}{k-2}(j+\frac12)^2}e^{2\pi i(\theta-w\tau) \alpha} }
{\vartheta_{11}
(\theta-w\tau,\tau+i\epsilon_1)} 
\sum_{n=-\infty}^{\infty}e^{2\pi i(\theta-w\tau) n}\cr\cr
&=&e^{i\pi ku}\frac{e^{2\pi i \tau\left(\frac{s^2}{k-2}+\frac{k}{4}w^2\right)}}
{\eta^3(\tau+i\epsilon_1)}  \sum_{m=-\infty}^{\infty}
e^{-2\pi im\left(\alpha+\frac k2 w\right)} 
\delta\left(\theta-w\tau+m\right)\, ,\label{chiwcon}
\eea
where the following identity was used
\bea
\sum_{n=-\infty}^{\infty}e^{2\pi ixn}=
\sum_{m=-\infty}^{\infty}\delta\left(x+m\right)\, .
\eea
In this case, the characters are defined as the weak limit 
$\epsilon_1,\,\epsilon_2^w\rightarrow0$, with the constraints
\bea
\left\{\begin{array}{lcr}
~~~~\epsilon_1>0\,,\cr
\epsilon_2^w-w\epsilon_1=0\, ,
\end{array}\right.
\eea
and they give  the
following contribution to the partition function:
\bea
Z_{\cal C}^{AdS_3}&=&
\sqrt{\frac{2-k}{8i(\tau_--\tau_+)}}\frac{e^{i\pi k(u_--u_+)}}
{\eta^3(\tau_-+i\epsilon_1)\eta^*{}^3(\tau_+-i\epsilon_1)}\cr
&&\times ~ \sum_{m,w=-\infty}^{\infty}e^{-2\pi
i\frac  k4 w(\theta_- -\theta_+)}\delta(\theta_- -w\tau_-+m)\delta(\theta_+
-w\tau_+ +m)\, ,~~\label{pfcon}
\eea
in agreement with the expression obtained in \cite{mo1}.

\subsection{Degenerate representations}

Degenerate representations  are not contained
 in the spectrum 
of the AdS$_3$ WZNW model but 
they  play an important role in the description of the boundary CFT. 
Indeed, using worldsheet duality, it was argued that they 
make up the Hilbert 
space of open string 
excitations of S$^2$ branes in the H$_3^+$ model \cite{GKS,PST}. 
For the analysis that we shall perform in the forthcoming sections, 
 it is useful to note the relation among their characters 
and those of 
discrete  and continuous representations of the universal cover 
of SL(2,${\mathbb R})$ discussed above.

The finite dimensional degenerate representations 
of SL(2,${\mathbb R}$) are labeled by the spin $j_{rs}^\pm$ defined 
by $1+2j^\pm_{rs}=
\pm \left(r+ s(k-2)\right)$, with $r,s+1=1,2,3,\dots$ for the 
upper sign 
and $r,s=1,2,3,\dots$ for the lower one. 
Here we consider 
$J=j^+_{r0}$, with characters  given by
\bea
\chi_J(\theta,\tau,u)=-\frac{2 e^{i\pi ku} e^{-2\pi i 
\tau\frac{(2J+1)^2}{4(k-2)}} 
\sin\left[\pi\theta (2J+1)\right]}{\vartheta_{11}(\theta+i\epsilon_2,\tau+i\epsilon_1)}\,,
\label{cdr}
\eea 
where the $\epsilon$'s are restricted to 
\bea
\left\{\begin{array}{lcr}
~\,\epsilon_1>0\,,\cr
|\epsilon_2|<\epsilon_1\,.
\end{array}\right.
\eea

Extrapolating the values of the 
spins in the expressions obtained in the previous sections, 
(\ref{cdr}) can be rewritten as 
\bea
\chi_J(\theta,\tau,u)
&=&\chi_{J}^{+,w=0}(\theta,\tau,u)+\chi_{-\frac k2-J}^{+,w=-1}(\theta,\tau,u)
-\chi_{J}^{\alpha=\{J\},w=0}(\theta,\tau,u)\,,\label{deg-dc}
\eea
where $\{J\}$ is the sawtooth function. 
Actually, this relation could have been guessed from a 
simple inspection of the spectrum (see Figure 1). This can be seen as a non trivial check of the characters defined above and,
 simultaneously, it shows the important role played by the $i0$ prescription in the definition of the characters of discrete representations. A naive computation of these characters, ignoring the $i0'$s, would yield the (wrong) conclusion 
$\chi_J=\chi^{+,w=0}_J+\chi^{+,w=-1}_{-\frac k2-J}$.

\centerline{\psfig{figure=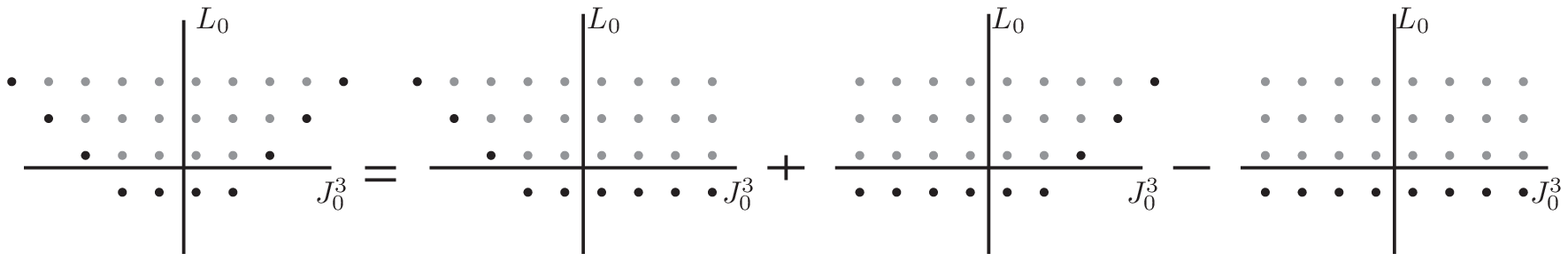,width=17.5cm}}\label{degesp}
{\footnotesize{Figure 1: The weight diagram of the degenerate representations
with spin $J=j^+_{r0}=\frac{r-1}2$, $r=1,2,3...$ can be decomposed as the
sum of the weight diagrams of the lowest and highest weight unflowed discrete
representations  minus that of the continuous representation
of spin $J$.}}
\section{Modular properties }\label{secMP}

The modular transformation $\tau\rightarrow\frac{a\tau + b}{c\tau + d}$, 
with integer 
parameters $a,b,c,d$ such that $ad-bc=1$, can be easily extended to include
$\theta, u$. Characters generating a representation space of the modular group
transform as \cite{dif}
\bea
\chi_{\mu} \left(\frac{\theta}{c\,\tau+d},\frac{a\,\tau+b}{c\,\tau+d},
u+\frac{c~\theta^2}{2(c\,\tau+d)}\right)=\sum_{\nu}M_{\mu}{}^{\nu} 
\chi_{\nu}(\theta,\tau,u),\label{modultrans}
\eea 
$M$ being the matrix associated to the group element.
Insofar as $\tau$ and $u$ are 
concerned, the sign of all the parameters $a,b,c,d$ may be
simultaneously changed without affecting the transformation. 
The modular group 
PSL(2, ${\mathbb Z})
=\frac{SL(2, {\mathbb Z})}{{\mathbb Z}_2}$ is
generated by 
$T=\left(\begin{matrix}1&1\cr 0&1\end{matrix}\right)$ and 
$S=\left(\begin{matrix}0&-1\cr 1&0\end{matrix}\right)$. These 
 transformations map
$\theta\rightarrow\theta$ and $\theta\rightarrow
\frac\theta\tau$, respectively, but  inverting the 
signs of $a,b,c,d$,  the mapping gives the opposite
sign for $\theta$.
Therefore,
the space spanned by the characters 
does not realize a good
representation space for the modular group unless 
the characters are symmetric under $\theta\leftrightarrow-\theta$,
$e.g.$ for self-conjugate representations.
When this is not the case,
 the characters form a representation of the double covering of the 
modular group, where $S^2$ is not the identity
but  the charge conjugation matrix.
In fact,
 $S^2$ produces time and parity inversion on the torus geometry and, by CPT
invariance, it transforms a character into its conjugate.

\subsection{The $S$ matrix} 

Below we will find
 explicit expressions for generalized $S$ transformations of the characters
 introduced in the previous section, 
setting $u=0$ for short, 
as\footnote{Some authors use the $\tilde S$ matrix generating
$\chi_{\mu}\left(-\frac{\theta}{\tau},-\frac{1}{\tau},u+\frac{\theta^2}
{2\tau}\right)$. This is given by $\tilde S_\mu{}^\nu = S_\mu {}^{\nu^ +}$,
 where $\nu ^+$ labels the conjugate $\nu$-representation.}
\bea
\chi_{\mu}(\frac{\theta}{\tau},-\frac{1}{\tau},0)=e^{-2\pi i 
\frac k4\frac{\theta^2}{\tau}} \sum_{\nu} S_{\mu}{}^{\nu}
\chi_{\mu}(\theta,\tau,0)\, ,\label{Stransf}
\eea
and we will show that, unlike standard expressions, 
they contain a sign of $\tau$ factor. This result can already
be inferred from the $S$ modular transformation of the partition 
function. Indeed,  ignoring the $\epsilon$'s and the
contact terms, one
 finds for the contributions from discrete representations\footnote{ 
$\tilde Z^{AdS_3}_{\mathcal D}$ is  the contribution to the partition 
function 
for $\theta$ and $\tau$  far from $\theta+n\tau\in{\mathbb Z}$, 
$\forall n\in{\mathbb Z}$.}
\bea
\tilde Z^{AdS_3}_{\cal D}
(\tau '_-,
\theta '_-,u '_- ;\tau '_+,
\theta '_+,u '_+)
~=~sgn(\tau_-\,\tau_+)~~
\tilde Z^{AdS_3}_{\cal D}
(\tau_-,\theta_-,u_-;\tau_+,\theta_+,u_+)\, ,
\ \ \ \ \  \ 
\label{ZDmod}
\eea
while the contributions from the continuous series verify
\bea
Z^{AdS_3}_{\cal C}(\tau '_-,
\theta '_-,u '_- ;\tau '_+,
\theta '_+,u '_+)~=~
 sgn(\tau_-\,\tau_+)~~
Z^{AdS_3}_{\cal C}(\tau_-,\theta_-,u_-;\tau_+,\theta_+,u_+) \, , \ \ \ \ \ \
\label{ZCmod}
\eea 
where the primes denote the $S$ modular transformed parameters.
This suggests that the block $S_{d_i}{}^{d_j}$, 
$d_i$ labeling discrete representations, is given by 
$sgn(\tau)~{\cal S}_{d_i}{}^{d_j}$ with
${\cal S}_{d_i}{}^{d_j}$ being unitary. Moreover, since  
 the characters of the continuous representations
 contain purely contact terms, 
one expects that 
 they close
among themselves. 
This together with (\ref{ZCmod}) suggest that the block $S_{c_i}{}^{c_j}$, 
$c_i$ labeling continuous representations, is given by 
$sgn(\tau)~{\cal S}_{c_i}{}^{c_j}$ with
${\cal S}_{c_i}{}^{c_j}$ being unitary. 
We will explicitly show these features of the generalized modular 
transformations in the next section. 
In this sense, the characters of the AdS$_3$ model on the Lorentzian torus
are pseudovectors with respect to the standard modular $S$ transformations.

A naive treatment of the Lorentzian partition function 
as a Wick rotation of the Euclidean path integral, would suggest
 the appearance of this sign after an $S$
transformation from the measure, when one
takes into account
the change in the metric  (see appendix A).
However, it will be clear from the results
of the next section, that the failure in the modular invariance of
$Z_{\cal D}^{AdS_3}$ is less
subtle than just  the sign appearing in (\ref{ZDmod}).

\subsubsection{Continuous representations }

The $S$ transformed characters of continuous representations can be written as:
\bea
\chi_j^{\alpha,w}(\frac{\theta}{\tau},-\frac{1}{\tau},0)=
\frac{e^{-2\pi i\left(\frac{s^2}{k-2}+\frac k4 w^2\right)\frac1\tau}}
{(-i\tau)^{\frac32}\eta^3(\tau+i\epsilon_1)} \sum_{m=-\infty}^{\infty} 
e^{2\pi im\alpha}\delta\left(\frac{\theta}{\tau}+\frac{w}{\tau}-m\right),
\eea
where $\eta(-\frac1\tau+i\epsilon_1)\equiv\eta(-\frac{1}{\tau+i\epsilon_1})=
e^{\mp \frac{i\pi}4}
\sqrt{|\tau |}~\eta(\tau+i\epsilon_1)$, the upper (lower) sign holding for 
$\tau>0$ ($\tau <0$).

Using
\bea
e^{-2\pi i\frac{s^2}{k-2}\frac1\tau}=e^{\mp\frac{i\pi}4}
\sqrt{\frac{2|\tau |}{k-2}}
\int_{-\infty}^{+\infty}ds'\; e^{-4\pi i\frac{ss'}{k-2}}
\; e^{2\pi i\tau \frac{s'{}^{2}}{k-2}}\, ,\label{tmodw0} 	
\eea
we find
\bea
\chi_j^{\alpha,w}(\frac{\theta}{\tau},-\frac{1}{\tau},0)&=&\frac{e^{-2\pi i
\frac k4\frac{\theta^2}{\tau}}}{\tau}\int_{-\infty}^{+\infty}ds'\; 
\tilde{\cal S}_s{}^{s'} \frac{e^{\frac{2\pi i}{k-2}\tau s'{}^{2}}}{{\eta^3(\tau+i\epsilon_1)} }\nonumber\\
&&\times~
\sum_{m=-\infty}^{\infty} e^{2\pi i\frac k4 \tau m^2} e^{2\pi im\alpha}~~
\delta\left(\frac{\theta}{\tau}+\frac{w}{\tau}-m\right)\, ,
\eea
with $~\displaystyle\tilde{\cal S}_s{}^{s'}=i\sqrt{\frac{2}
{k-2}}e^{-4\pi i\frac{ss'}{k-2}}$.

From $\delta\left(\frac{\theta}{\tau}+\frac{w}{\tau}-m\right)=
|\tau|~ \delta\left(\theta+w-m\tau\right)$ and renaming variables, one gets
\bea
\chi_j^{\alpha,w}(\frac{\theta}{\tau},-\frac{1}{\tau},0)&=&e^{-2\pi i
\frac k4\frac{\theta^2}{\tau}}sgn(\tau)\nonumber\\
&\times &\sum_{w'=-\infty}^{\infty} 
\int_{-\infty}^{+\infty}ds' \tilde{\cal S}_s{}^{s'} \frac{e^{2\pi i\tau
\left(\frac{ s'^{2}}{k-2}+\frac k4 w'^{2}\right)}}{\eta^3(\tau+i\epsilon_1)}  
e^{2\pi iw'\alpha}\delta\left(\theta-w'\tau+w\right) .
\label{chiconttrans}
\eea
In order to reconstruct the character  $\chi_{j'}^{\alpha',w'}$ 
in the $r.h.s.$, we use the identity
\bea
\delta\left(\theta-w'\tau+w\right)=
\sum_{m'=-\infty}^\infty\int_0^1 d\alpha' e^{2\pi i \left(w\alpha'+\frac k2 
ww'\right)}
e^{-2\pi im'\left(\alpha'+\frac k2 w'\right)}
\delta\left(\theta-w'\tau+m'\right)\, ,\label{conttrick}
\eea
and exchanging summation and integration\footnote{Here, summation and 
integration can be exchanged because, for a fixed $w'$,
 the series always reduces to a finite sum when it is considered
as a distribution acting on a test function.}, 
(\ref{chiconttrans}) can be rewritten as
\bea
\chi_j^{\alpha,w}(\frac{\theta}{\tau},-\frac{1}{\tau},0)=e^{-2\pi i
\frac k4\frac{\theta^2}{\tau}}sgn(\tau)\sum_{w'=-\infty}^{\infty} 
\int_{0}^{+\infty}ds'\; \int_0^1 d\alpha'~~
{\cal S}_{s,\alpha,w}{}^{s',\alpha',w'} \chi_{j'=-\frac12+is'}^{\alpha',w'}
(\theta,\tau,0)\, ,\nonumber
\eea
with
\bea
{\cal S}_{s,\alpha,w}{}^{s',\alpha',w'}= 2i\sqrt{\frac{2}{k-2}}~~
\cos\left(4\pi 
\frac{ss'}{k-2}\right)~~ e^{2\pi i \left(w\alpha'+w'\alpha+\frac k2
  ww'\right)}\, ,
\eea
which is symmetric and, 
as expected from (\ref{ZCmod}), unitary, $i.e.$
\bea
\sum_{w'=-\infty}^{\infty}\int_{0}^\infty ds'\int_0^1 d\alpha' {\cal S}_{s_1,
\alpha_1,w_1}{}^{s',\alpha',w'}
{\cal S}^{\dagger}_{s',\alpha',w'}{}^{s_2,\alpha_2,w_2}=\delta(s_1-s_2)
\delta(\alpha_1-\alpha_2)\delta_{w_1,w_2}\, .
\eea

\subsubsection{Discrete representations}\label{secSdiscete}

The structure of the characters of the discrete representations
 is more involved than that of the continuous ones. A priori, we expect
 that characters of
both discrete and continuous representations
appear in the generalized modular transformations. So, generically we can
 assume
\bea
\chi_j^{+,w}(\frac\theta\tau,-\frac1\tau,0)&=&
e^{-2\pi i
\frac k4\frac{\theta^2}{\tau}}sgn(\tau) \sum_{w'=-\infty}^{\infty}\left\{ 
\int_{-\frac{k-1}{2}}^{-\frac12}{\cal S}_{j,w}{}^{j',w'}~\chi_{j'}^{+,w'}(\theta,\tau,0)\right.\cr\cr
&& ~~~~~~\qquad ~~~~~~~~
+~
\left.\int_0^1 d\alpha'\int_0^\infty ds'~{\cal S}_{j,w}{}^{s',\alpha',w'}~\chi_{j'=-\frac12+is'}^{\alpha',w'}(\theta,\tau,0) \right\}\, .\nonumber
\eea
Fortunately, it is easy to separate the contributions from
 discrete and continuous representations. If one considers generic
values of $\theta$ and $\tau$ far from $\theta+n\tau\in{\mathbb Z}$ 
for $n\in{\mathbb Z}$, the contributions of the continuous series
 in the $r.h.s.$ can be neglected as well as all contact terms and 
$\epsilon$'s. On the other hand, if 
$\theta+n\tau\notin{\mathbb Z},\forall n\in {\mathbb Z}$ then
 $\frac\theta\tau-p\frac1\tau\notin{\mathbb Z},\forall p\in {\mathbb Z}$ 
and  all contact terms and $\epsilon$'s of the $l.h.s.$ can be neglected 
too. Thus, we obtain
\bea
\chi_j^{+,w}(\frac\theta\tau,-\frac1\tau,0)&=&
\frac{(-)^we^{\frac{2\pi i}
{k-2}\frac1\tau \left(j+\frac12-w\frac{k-2}{2}\right)^2}e^{-2\pi i\frac\theta
\tau \left(j+\frac12 -w\frac{k-2}{2}\right)}} {i\vartheta_{11}(\frac\theta
\tau,-\frac1\tau)}\nonumber\\
&=&(-)^{w+1}
\frac{
e^{\frac{2\pi i}{k-2}\frac1\tau \left(j+\frac12-(w+\theta)
\frac{k-2}{2}\right)^2}
e^{-2\pi i\frac k4\frac{\theta^2}{\tau}}} 
{i\sqrt{i\tau}\vartheta_{11}(\theta,\tau)}\, ,\label{tmd}
\eea
where the following identity was  used for 
$\tau\in \mathbb R$:
 \bea
\vartheta_{11}(\frac\theta\tau,-\frac1\tau)=
\mp e^{\pi i \frac{\theta^2}{\tau}}e^{\pm i\frac\pi4 }\sqrt{|\tau|}\;
\vartheta_{11}(\theta,\tau)\, ,\label{thetataureal}
\eea
the upper (lower) sign holding for
$\tau>0$ ($\tau<0$ ).
Inserting
\bea
e^{\frac{2\pi i}{k-2}\frac1\tau \left(j+\frac12-(w+\theta)\frac{k-2}{2}
\right)^2}= e^{\pm i\frac\pi4}\sqrt{\frac{2|\tau|}{k-2}}
\int_{-\infty}^{+\infty}
d\lambda'e^{\frac{4\pi i}{k-2} 
\lambda'\left(j+\frac12-(w+\theta)\frac{k-2}{2}\right)} 
e^{-\frac{2\pi i}{k-2}\tau\lambda^{'2}} \label{gaussian}
\eea
into (\ref{tmd}), changing the integration variable 
to $j'+\frac 12-w'\frac {k-2}2$ 
and using (\ref{thetataureal}), we get
\bea
\chi_j^{+,w}(\frac\theta\tau,-\frac1\tau,0)=e^{-2\pi i
\frac k4\frac{\theta^2}{\tau}} sgn(\tau)
\sum_{w'=-\infty}^{\infty}\int^{-\frac{1}{2}}_{-\frac{k-1}{2}}dj' 
{\cal S}_{j,w}{}^{j',w'}\chi_{j'}^{+,w'}(\theta,\tau,0), \label{chidis}
\eea
with
\bea
{\cal S}_{j,w}{}^{j',w'}=(-)^{w+w'+1}~\sqrt{\frac{2}{k-2}}~e^{\frac{4\pi i}
{k-2} 
\left(j'+\frac12-w'\frac{k-2}{2}\right)\left(j+\frac12-w\frac{k-2}{2}\right)}
\, .
\label{SD}
\eea
Notice that this block of the  ${\cal S}$ matrix 
is symmetric and, again as expected 
from (\ref{ZDmod}), unitary\footnote{Changing 
$e^{\pm i\frac\pi4}\sqrt{\tau}$ by $\sqrt{i\tau}$, the validity of
 (\ref{gaussian}) can be extended to the full 
lower half plane and that of
(\ref{thetataureal}) can be extended to the upper half 
plane, giving 
\bea
\vartheta_{11}(\frac\theta\tau,-\frac1\tau)=
- e^{\pi i \frac{\theta^2}{\tau}}\sqrt{i\tau}\;
\vartheta_{11}(\theta,\tau)\, .\label{stheta}
\eea
If one naively cancels the $\sqrt{i\tau}$ 
terms and ignores the sign factor due to the
different branches, a $\tau$ independent expression is obtained
for the $S$ matrix. However, such
$S$ matrix does not
obey the properties $S^2=(ST)^3= C$,  
$C$ being the charge 
conjugation matrix, but the opposite ones.}.

While the identity (\ref{gaussian}), which 
is essential to reconstruct the discrete characters in the
$r.h.s.$ of (\ref{chidis}), only makes sense 
for Im $\tau\leq0$, the characters are only
well defined  for Im $\tau\geq0$. Therefore, to determine 
 the generalized $S$ transformation, it is  crucial 
that $\tau\in {\mathbb R}$. 

\medskip

Finding the block 
${\mathcal S}_{j,w}{}^{s',\alpha',w'}$ 
mixing discrete with continuous representations 
is a much more technical issue, which we discuss in appendix B. 
Here we simply display the result, namely
\bea
{\cal S}_{j,w}{}^{s',\alpha',w'}=-i\sqrt{\frac{2}{k-2}}e^{-2\pi i
\left(w'j-w\alpha'-ww'\frac k2\right)} \left[\frac{e^{\frac{4\pi}{k-2}s'
\left(j+\frac12\right)}}{1+e^{-2\pi i(\alpha'-is')}} + 
\frac{e^{-\frac{4\pi}{k-2}s'\left(j+\frac12\right)}}
{1+e^{-2\pi i(\alpha'+is')}}\right]. \label{degcont}~~~~
\eea

This block prevents the full ${\mathcal S}$ matrix 
from being unitary. Instead, we find ${\mathcal S}^*{\mathcal S}=id$. 
This implies that the full partition function defined from the product
of characters is not modular invariant, not only due to the 
sign of the modular 
parameters. Actually,
after a modular transformation, the mixing block introduces terms where the 
left modes are in discrete representations
and the right ones in continuous series,
and vice versa, as well as new terms containing left and right continuous 
representations.

In section 
\ref{STproperties}, we explicitly check that the blocks of the 
$S$ matrix determined here have the correct properties.

\subsubsection{Degenerate representations} 
\label{Sdeg}

The modular properties discussed 
above can be used to write the 
$S$ transformation of the characters of the
degenerate representations 
with $1+2J\in{\mathbb N}$
as:
\bea
\chi_J(\frac\theta\tau,-\frac1\tau,0)&=&e^{-2\pi i\frac k4 \frac{\theta^2}
{\tau}}
sgn(\tau)\sum_{w=-\infty}^\infty \left\{\int^{-\frac12}_{-\frac{k-1}{2}} dj~
{\cal S}_J{}^{j,w}\chi_j^{+,w}(\theta,\tau,0)\right.\cr 
&&~~~~~~~~~~~~~~~~~~~~~~~~~~
+~\left.\int_0^1 d\alpha\int_{-\frac{k-1}{2}}^{-\frac12} ds ~
{\cal S}_{J}{}^{s,\alpha,w}\chi_{j=-\frac12+is}^{\alpha,w}(\theta,\tau,0)
\right\},\nonumber
\eea
where
\bea
{\cal S}_J{}^{j,w}=2i\sqrt{\frac{2}{k-2}}(-)^{w+1} \sin
\left[\frac{\pi}{k-2}\left(1+2j-w
(k-2) \right)\left(2J+1\right)\right]\,,
\eea
and
\bea
{\cal S}_J{}^{s,\alpha,w}&=&-i(-)^{2Jw}\sqrt{\frac{2}{k-2}}e^{\frac{4\pi}
{k-2}s(J+\frac 12)}\left(1+\frac{1}{1+e^{-2\pi i (\alpha-is)}}+\frac{1}{1+
e^{2\pi i (\alpha+is)}}\right) \nonumber\\
&&+~~(s\leftrightarrow -s) \, .
\label{degs}
\eea

\subsection{The $T$ matrix }
Together with the $S$ matrix, the  $T$ matrix defines a basis over the space 
of modular transformations.
Using
\bea
\vartheta_{11}(\theta,\tau+1)=e^{\frac{\pi i}{4}}\vartheta_{11}(\theta,\tau),
\qquad
\eta(\tau+1)=e^{\frac{\pi i}{12}}\eta(\tau)\, ,
\eea
the characters of the 
discrete and continuous representations transform respectively with 
\bea
T_{j,w}{}^{j',w'}=\delta_{w,w'}\delta(j-j')e^{-\frac{2\pi i}{k-2}
\left(j'+\frac12-w'\frac{k-2}{2}\right)^2-\frac{\pi i}{4}}
\eea
and 
\bea
T_{s,\alpha,w}{}^{s',\alpha',w'}=\delta_{w,w'}\delta(\alpha-\alpha')
\delta(s-s')e^{2\pi i\left(\frac{ s^2}{k-2}-\frac{k}{4}w^2-w\alpha-
\frac 18\right)}\, ,
\eea
while the $T$
 transformation of the characters of the degenerate representations is given by
\bea
\chi_J(\theta,\tau+1,0)=e^{-\frac{2\pi i}{k-2}\left(J+\frac12\right)^2}
e^{-\frac{\pi i}{4}}\chi_J(\theta,\tau,0)\, .
\eea

\subsection{Properties of the $S$ and $T$ matrices}\label{STproperties}
 
The expressions 
$(ST)^3$ and $S^2$ must give the conjugation matrix, $C$.
We have found above
that the characters of the AdS$_3$  model do not expand a 
representation space for the modular group since the generators depend on the
sign of
$\tau$. Nevertheless, in terms of the $\tau$ independent part of $S$, 
that we have denoted ${\mathcal S}$, these identities read $C=
(ST)^3=sgn(\tau+1)sgn(\frac{\tau}{\tau+1})sgn(-\frac1\tau)({\mathcal S}T)^3=-({\mathcal S}T)^3$ and $C=S^2=sgn(\tau)sgn(-\frac1\tau){\mathcal S}^2=-{\mathcal S}^2$.

As a consistency check on the  expressions found above for $S$ and $T$,
 an explicit computation gives 
\bea
-({\cal S}T)^3{}_{j_1,w_1}{}^{j_2,w_2}=-{\cal S}^2{}_{j_1,w_1}{}^{j_2,w_2}= 
\delta_{w_1+w_2+1,\,0}~\delta\left(j_1+j_2+\frac k2\right)\, ,
\eea
which corresponds to the conjugation matrix restricted to the discrete 
sector, since
$\hat {\mathcal D}_j^{+,w}$ is the conjugate representation of
$\hat {\mathcal D}_j^{-,-w}$, which in turn can be identified with
$\hat {\mathcal D}_{-\frac k2-j}^{+,-w-1}$ using the spectral flow symmetry.
Similarly,
for the block of continuous representations we get
\bea
-({\cal S}T)^3_{s_1,\alpha_1,w_1}{}^{s_2,\alpha_2,w_2}=
-{\cal S}^2{}_{s_1,\alpha_1,w_1}{}^{s_2,\alpha_2,w_2}= 
\delta_{w_1,-w_2}\delta(s_1-s_2)\delta(\alpha_1+\alpha_2-1)\, ,\label{STcontprop}
\eea
which is
again the charge conjugation matrix, since $\hat {\mathcal C}_j^{1-\alpha,-w}$
is the conjugate representation of
$\hat{\mathcal C}_j^{\alpha,w}$. 

Of course, one also needs to show that the non diagonal terms vanish. 
The equalities
$({\mathcal S}T)^3{}_{s_1,\alpha_1,w_1}{}^{j_2,w_2}=
{\mathcal S}^2{}_{s_1,\alpha_1,w_1}{}^{j_2,w_2}=0$ are trivially 
satisfied as a consequence of 
${\mathcal S}_{s_1,\alpha_1,w_1}{}^{j_2,w_2}=0$. One can also show that
$({\mathcal S}T)^3{}_{j_1,w_1}{}^{s_2,\alpha_2,w_2}=
{\mathcal S}^2{}_{j_1,w_1}{}^{s_2,\alpha_2,w_2}$=0, but this computation
is more involved, so
the details are left to appendix \ref{MixBlockApp}.

\section{Revisiting D-branes in AdS$_3$}\label{Dbranes}

D-branes can be characterized by the one-point
functions of the states in the
bulk, 
living on the upper half plane.
In RCFT, these one-point functions
can be determined from the entries of the $S$ matrix, a property 
that we will call a {\it Cardy structure}. This property is closely 
related to the Verlinde formula and, {\it a priori},
there is no reason for it to hold in non RCFT.
In this section we explore this relation
 in the  AdS$_3$ model.

D-branes in AdS$_3$ and related models 
have been studied in several works (see for 
instance \cite{GKS}-\cite{AS} 
and references therein). 
Because 
 the Lorentzian AdS$_3$ geometry is obtained by sewing 
an infinite number of  SL(2,${\mathbb R})$ group manifolds,
 the corresponding D-brane solutions 
 are trivially obtained from those of SL(2,${\mathbb R})$.
The geometry of these D-branes  was considered 
semiclassically in \cite{BP}, where it was found
 that solutions of the Dirac Born Infeld  action stand for regular and
 twined 
conjugacy classes of SL(2,${\mathbb R})$. 

Here, we shall restrict to the maximally symmetric D-branes 
discussed in 
\cite{BP}. The model also has symmetry breaking D-brane solutions,
 but in this case, the open string 
spectrum is not a sum of SL(2,${\mathbb R})$ representations and 
then we do not expect the one-point functions to be determined by
the $S$ matrix.
We begin this section with a short introduction to the geometry of  D-branes 
in AdS$_3$. 
A very 
comprehensive study about the (twined) conjugacy classes of 
SL(2,${\mathbb R})$ and a semiclassical analysis 
of branes can be found in \cite{Stanciu} and \cite{BP}.
 Both can be easily 
extended to the universal covering. 
Here, we 
review the analysis of the 
conjugacy classes in order to make the discussion self contained
and discuss the extension to the universal covering. 

Then we turn to the explicit construction of the Ishibashi 
states for  regular and twisted boundary gluing conditions 
which give rise to the maximally symmetric 
D-branes. These equations were solved in the past for
 the single cover of SL(2,${\mathbb R})$ (see \cite{RR} for
 twisted gluing conditions) 
with different amounts of spectral flow in the left and right 
sectors, namely $w_L=-w_R$, and 
therefore, these solutions are not contained in
 the spectrum of the AdS$_3$ model 
(with 
the obvious exception of 
$w=0$ discrete
and $w=0$, $\alpha=0,\frac12$  continuous representations).

We find that the one-point functions
 of states in discrete representations coupled 
to point-like and H$_2$ branes exhibit a
{\it Cardy structure} and we propose a generalized 
Verlinde formula giving the fusion rules of the
 degenerate representations with $1+2J\in {\mathbb N}$. 

\subsection{Conjugacy classes in AdS$_3$}\label{conjclass}

Elements of SL(2,${\mathbb R})$ can be parametrized by four real parameters 
$X_0,\dots,X_3$ as
\bea
g=\frac{1}{\ell}\left(\begin{matrix}X_0+X_1 & X_2+X_3\cr
                    X_2-X_3 & X_0-X_1\end{matrix}\right)\, ,
\eea
with $X_0^2-X_1^2-X_2^2+X_3^2=\ell^2$. This gives a representation of 
the SL(2, ${\mathbb R})$ group manifold 
embedded in a 4 dimensional flat space.
 When the signature of this embedding space is  $(-1,1,1,-1)$,
it corresponds to a pseudosphere whose covering space is AdS$_3$. 

A more convenient coordinate system is given by
\bea
X_0+iX_3&=&\ell e^{it}\cosh\rho\, ,\cr
X_1+iX_2&=&\ell e^{i\theta}\,\sinh\rho\, ,
\eea     
where AdS$_3$ is simply obtained by decompactifying the 
timelike direction $t$.  

As is well known \cite{AS}, the world-volume of a 
symmetric D-brane on the SL(2,${\mathbb R})$ group manifold is 
given by the (twined) conjugacy classes 
\bea
{\cal W}_g^\omega=\left\{\omega
(h)gh^{-1},\forall h\in SL(2,{\mathbb R})\right\}\, ,
\eea
where $\omega$ determines the gluing condition 
connecting left and right moving currents,  
$\omega(g)=\omega^{-1}g\,\omega$. When $\omega$ 
is an inner automorphism, ${\mathcal W}_g^\omega$ can be seen as left 
group translations of the 
regular conjugacy class (of the element $\omega g$). So, one can restrict 
attention to the case 
$\omega=id.$, and the conjugacy classes 
are simply given by the solution to
\bea
tr~ g=2\frac{X_0}{\ell}=2\tilde C\, .
\eea  

The 
geometry of the world-volume is then parametrized by the constant $\tilde C$ as
\bea
-X_1{}^2-X_2{}^2+X_3{}^2=\ell^2\left(1-\tilde C^2\right)\, .
\eea
Different geometries can be distinguished for $\tilde C^2$  bigger,
 equal or smaller than one. 
The former gives rise to a two dimensional de Sitter space, dS$_2$,
the latter to a two dimensional hyperbolic 
space, H$_2$, and the case $|\tilde C|=1$ splits into
 three different geometries: the apex,
the future and the past of a light-cone.

 A more convenient way to parametrize these solutions is given 
by the redefinition   
\bea
\tilde C= \cos\sigma\, .
\eea   

For $|\tilde C|>1$, $\sigma=ir+\pi v,~r\in{\mathbb R^+},~v\in{\mathbb Z_2}$.
The world-volumes are given by 
\bea
\cosh\rho \cos t=\pm\cosh r\, .
\eea
Each circular D-string is emitted and absorbed at the boundary in a time 
interval of width $\pi$ but does not reach the origin 
unless $r=0$. Their lifetime is determined by $v$.
 
For $|\tilde C|<1$, $\sigma$ is real and   
\bea
\cosh\rho \cos t=\cos \sigma\, .
\eea
If one restricts $\sigma\in(0,\pi)$, there are two different solutions 
 for each $\sigma$, for instance 
one with $t\in(-\frac\pi2,-\sigma]$ and another one
with $t\in[\sigma,\frac{3\pi}{2})$. To distinguish between these two 
solutions we 
can take
$\sigma=\lambda+\pi v,~\lambda\in\left(-\pi,0\right),~v\in{\mathbb Z_2}$, 
 such that $t={\rm arcos}(\cos\sigma/\cosh\rho)$, taking the branch 
where $t=\sigma$ when it crosses over the origin.
Because these solutions have Euclidean signature, they are identified 
as instantons in AdS$_3$. In fact, they represent constant time slices in hyperbolic coordinates.

For $|\tilde C|=1$, $\sigma=0$ or $\pi$ and
\bea
\cosh\rho\cos t=\pm1\, .
\eea
For example, for 
$\tilde C=1$, this corresponds to a circular D-string at the boundary at
$t=-\pi/2$ collapsing 
to the instantonic solution in $\rho=0$ at $t=0$, and
 then expanding again to a D-string reaching the boundary at $t=\pi/2$.

All of these solutions are restricted to the single covering of 
SL(2,${\mathbb R}$). In the universal covering, $t$ is decompactified 
and  the picture is periodically repeated. 
The general solutions can be parametrized by a pair $(\sigma,q)$, 
$q\in{\mathbb Z}$, or equivalently, the range of $\sigma$ can be extended to
 $\sigma=ir+q\pi$ for dS$_2$ branes, 
$\sigma=\lambda+q\pi$ for H$_2$ branes or $\sigma=q\pi$
 for point-like and light-cone branes.

Preparing for
the discussions on  one-point functions and  $Cardy~structure$, it 
is interesting to note that these parameters can be naturally identified 
with representations 
of the model. For instance, one can label the 
D-brane solutions 
as
\bea
\sigma=\frac{2\pi}{k-2}\left(j+\frac12-w\frac{k-2}{2}\right)
\, ,\label{pardS2} 
\eea
with $j=-\frac12+is,
~s\in{\mathbb R^+},~w\in{\mathbb Z}$ for dS$_2$ branes, 
$j\in\left(-\frac{k-1}{2},-\frac12\right),~w\in{\mathbb Z}$ for H$_2$ 
branes   
and finally $\sigma=n\pi,~n\in{\mathbb Z}$ for
 the point-like and light-cone D-brane solutions.

The appearance of
the level $k$ in a classical regime could seem awkward.
However, it is useful to  recall
that $\sigma$ is just a parameter labeling the conjugacy classes, and
the factor $k-2$ can be eliminated  by simply redefining 
 $j$ through a change of variables. 
The important observation is that this suggests 
 $\sigma$ labels the exact solutions, $e.g$ the one-point 
functions at finite $k$ will be found to be parametrized
exactly by (\ref{pardS2}) and in fact, in the semiclassical 
regime $k\rightarrow\infty$, the domain of $\sigma$ does not change at all.
\bigskip

When $\omega$ is an outer automorphism, 
one can take 
$\omega=\left(\begin{matrix}0&1\cr1&0\end{matrix}\right)$
up to group translations.  
In this case, the twined conjugacy classes are given by
\bea
tr~ \omega g=2\frac{X_2}{\ell}=2C\, .
\eea
The world-volume geometry now describes an AdS$_2$ space for all $C$ since
\bea
X_0^2-X_1^2+X_3^2=\ell^2(1+C^2)\, .
\eea
These are static open D-strings with endpoints fixed at the boundary. This is obvious in cylindrical coordinates, $i.e.$
\begin{equation}
\sinh\rho\sin\theta=\sinh r\, ,
\end{equation}
where we have renamed $C=\sinh r$. So, 
after decompactifying the time-like direction $t$, there is no 
need to extend the domain of 
$r$. 

Let us end this brief review with a word of caution. 
In this section we have reviewed the 
twined conjugacy classes and, although branes wrap conjugacy classes, 
extra restrictions appear when studying the semiclassical or 
exact solutions. In particular,
it was found  in \cite{BP} that 
 $r$ becomes a positive quantized parameter at the semiclassical
 level.

\subsection{Coherent states}

Boundary states play a fundamental role in understanding boundary conformal 
field theories. They 
store all the information about possible D-brane solutions and their 
couplings to bulk states. 
Even though there is no  systematic method to obtain all possible boundary 
states of an arbitrary model, if one 
works in the boundary theory of a given WZNW model and  looks for 
special D-brane configurations with 
more symmetries than the conformal one, $e.g.$ the symmetry generated by a
 given subalgebra of the original 
current algebra, then the procedure is more tractable because these 
symmetries impose extra restrictions, which together 
with certain {\it sewing constraints}, can be used to obtain exact solutions. 
Following these ideas, one can study different gluing conditions for the 
left and right current modes, 
consistent with the 
affine algebra \cite{KO} as well as with 
the conformal symmetry via the Sugawara construction \cite{Ishibashi}.

In the case of AdS$_3$, much of the progress reached
in this direction is based on the analytic continuation from 
H$_3^+$ \cite{PST}.
Gluing conditions were imposed
as differential equations 
applied directly to find, with 
the help of certain {\it sewing contraints}, the one-point 
functions of maximally symmetric D-branes. It would 
be interesting to get the one-point functions of the AdS$_3$ 
model without reference to other models, but the approach used so far
 cannot 
be easily extended. In the first place, it was developed in
 the $x$-basis  of the H$_3^+$ model, which is not a good 
basis for the representations of the universal 
covering of SL(2,${\mathbb R})$. 
 Suitable bases instead
are the $m$-
or $t$-basis  \cite{mo1, rib}. Moreover,
 there are still some open questions about the fusion rules of the AdS$_3$
model \cite{wc} 
which deserve further attention before analyzing the 
{\it sewing constraints}. Therefore, we will not compute the one-point 
functions in this way, but will give the first step in this 
direction by finding the explicit expressions for the Ishibashi states in 
the $m$-basis for all 
the representations of the Hilbert space of the bulk theory.

\subsubsection{Coherent states for regular gluing conditions} 
\label{coherentstates}             

Boundary states associated to dS$_2$, H$_2,$ light-cone and point-like
 D-branes in AdS$_3$ must satisfy the following regular gluing conditions 
\cite{RR}
\bea
&&\left(J_n^3-\bar J_{-n}^3\right)\left|{\bf s}\right\rangle=0,\cr
&&\left(J_n^{\pm}+\bar J_{-n}^{\mp}\right)\left|{\bf s}\right\rangle=0,
\label{boundstatecond}
\eea
where ${\bf s}$ labels the members of the family of branes 
allowed by the gluing conditions.

These constraints are linear and leave  each representation invariant, 
so that the boundary states
 must be expanded as a sum of solutions in each module. 
The solutions represent coherent states, usually called
 Ishibashi states \cite{Ishibashi}. 

Let us begin introducing the following notation
which will be useful in the 
subsequent discussions. Let
\bea
\left|j,w,\alpha;n, m\right\rangle=\left|j,w,\alpha\right\rangle
\{\left|n\right\rangle\otimes\overline{\left| m\right\rangle}\},~~
\left|j,w,+;n, m\right\rangle=\left|j,w,+\right\rangle\{\left|n\right
\rangle\otimes\overline{\left| m\right\rangle}\},
\eea
denote  orthonormal bases for 
$\hat{\mathcal C}_j^{\alpha,w}\otimes\hat{\mathcal C}_j^{\alpha,w}$ 
and $\hat{\mathcal D}_j^{+,w}\otimes\hat{\mathcal D}_j^{+,w}$, 
respectively. They 
satisfy\footnote{The separation between 
$|j,w,\alpha\rangle$ or $|j,w,+\rangle$ 
and $|n\rangle, \overline{|m\rangle}$
 in different kets is simply a matter of 
 useful notation for calculus and does not denote tensor product.}
\bea
\langle j,w,\alpha;n, m|j',w',\alpha';n',m'\rangle &=&~~~~~
\langle j,w,\alpha|j,w,\alpha\rangle~\times ~ 
\langle n|n'\rangle\times \overline{\langle m}
|\overline{m'\rangle}\cr
&=&\delta(s-s')\delta_{w,w'}\delta(\alpha-\alpha')~\epsilon_n\delta_{n,n'}~
\epsilon_m \delta_{m,m'},\cr\cr
\left\langle j,w,+;n,m|j',w',+;n',m'\right\rangle &=&\left\langle j,w,+
|j,w,+\right\rangle \left\langle n|n'\right\rangle \overline{\left
\langle m\right.}|\overline{\left.m'\right\rangle}\cr
&=&~ ~  \delta(j-j')\delta_{w,w'}~\epsilon_n\delta_{n,n'}~\epsilon_m 
\delta_{m,m'},
\eea
$\left\{|n>\right\}$ is an orthonormal basis in $\hat{\mathcal C}_j^{\alpha,w}$ 
(or $\hat{\mathcal D}_j^{+,w}$) for which the expectation values of 
$J^3_n, J^{\pm}_n$ are real numbers and $\epsilon_n=\pm1$ is its 
norm squared.
It is constructed by the action of the affine currents over the ket 
$|j,m=\alpha,w>=U_w|j,m=\alpha>$  ($|j,m=-j,w>$).

The Ishibashi states for continuous and discrete representations are found to be 
\bea
\left|j,w,\alpha\right.\gg= \sum_{n} \epsilon_n\bar V \left|j,w,\alpha;n,n
\right\rangle~~~ 
{\rm and}\quad ~
\left|j,w,+\right.\gg= \sum_{n} \epsilon_n\bar V\left|j,w,+;n,n\right\rangle,~~~\label{Ishdef}
\eea
respectively, where $V$ is defined as the linear operator satisfying 
\bea
V\prod_I J_{n_I}^{a_I}\left|j,m=-j,w\right\rangle &=& \prod_I \eta_{a_I b_I}
J^{b_I}_{n_I}\left|j,m=-j,w\right\rangle,\cr
V\prod_I J_{n_I}^{a_I}\left|j,m=\alpha,w\right\rangle & =& \prod_I 
\eta_{a_I b_I}J^{b_I}_{n_I}\left|j,m=\alpha,w\right\rangle,\label{Vdef}
\eea 
with $a=1,2,3$, $\eta_{ab}=diag(-1,-1,1)$
and the bar  denotes 
 action restricted to the antiholomorphic 
sector. It is easy to see that this defines a	unitary operator. 
The proof that 
they are solutions to (\ref{boundstatecond}) follows similar lines as 
those of \cite{Ishibashi}. 
As an example, let us consider an arbitrary base state 
$|j',w',\alpha';n',m'>$:
\bea
&&<j',w',\alpha';n',m'|J_r^3-\bar J_{-r}^3|j,\alpha,w\gg =\cr
&& ~~~
\delta(s-s')\delta_{w,w'}\delta(\alpha-\alpha')\sum_{n}\epsilon_n 
\left\langle n'\right|J_n^3\left|n\right\rangle\overline{\left\langle 
m'\right|}\bar V\overline{\left|n\right\rangle}-\epsilon_n 
\left\langle n'\right|\left.n\right\rangle \overline{\left\langle m'
\right|}\bar J_{-n}^3 \bar V\overline{\left|n\right\rangle}=\cr
&&  ~
\delta(s-s')\delta_{w,w'}\delta(\alpha-\alpha')\sum_{n}\epsilon_n 
\left\langle n'\right|J_n^3\left|n \right\rangle \left\langle n\right|
V\left|m'\right\rangle-\epsilon_n \left\langle n'\right|\left.n
\right\rangle \left\langle n\right|VJ_{n}^3 \left|m'\right\rangle=0\, .
\nonumber
\eea

The normalization fixed above for the Ishibashi states implies
\bea
\ll j,w,\alpha|e^{\pi i\tau\left(L_0+\bar L_0-\frac{c}{12}\right)}
e^{\pi i\theta(J_0^3+\bar J_0^3)}|j',w',\alpha'\gg &=& \delta (s-s')
~\delta_{w,w'}~\delta(\alpha-\alpha')~\chi_j^{\alpha,w}(\tau,\theta),\cr\cr
\ll j,w,+|e^{\pi i\tau\left(L_0+\bar L_0-\frac{c}{12}\right)}
e^{\pi i\theta(J_0^3+\bar J_0^3)}|j',w',+\gg &=&
 \delta(j-j')~\delta_{w,w'}~\chi_j^{+,w}(\tau,\theta).\label{IshNorm}
\eea

\subsubsection{{\it Cardy structure}
 and one-point functions for point-like branes}
\label{pointbranes}

Assuming that after Wick rotation the open string partition 
function in AdS$_3$ reproduces that of the H$_3^+$ model and a generalized 
Verlinde formula, we show in this section that 
 the one-point 
functions on
 localized branes in AdS$_3$ previously  found 
in \cite{israel}
can be recovered.
We also verify that the  one-point 
functions on  point-like and H$_2$ D-branes
exhibit a {\it Cardy structure}. 
Usually, this structure is accompanied by a Verlinde formula for 
the representations appearing in the boundary
 spectrum. In fact, the {\it Cardy structure} is a natural 
solution to the {\it Cardy condition} when the
Verlinde theorem holds. However, as we shall discuss, the latter 
does not hold in the AdS$_3$ WZNW model. 
The generalized 
Verlinde formula proposed in  appendix C  
reproduces the fusion rules of the degenerate representations, but it gives
contributions to the fusion rules of the discrete representations with 
an arbitrary amount of spectral flow, thus contradicting the 
selection rules determined in \cite{mo3}. Nevertheless, we find a
{\it Cardy structure}. 
\bigskip

{\textbf{\textit{Boundary states}}}
\medskip

Worldsheet duality allows to write the one loop partition function 
for open strings ending on point-like branes labeled by ${\bf s}_1$ and 
${\bf s}_2$ as
\bea
e^{-2\pi i\frac k4\frac{\theta^2}{\tau}}~Z_{{\bf s}_1{\bf s}_2}^{AdS_3}(\theta,\tau,0) &=&
\left
\langle \left.\Theta {\bf s}_1\right|\tilde q^{H^{(P)}}\tilde z^{J_0^3}
 \left|{\bf s}_2
\right.\right\rangle ~~\cr &=& \sum_{w=-\infty}^{\infty} 
\int^{-\frac12}_{-\frac{k-1}{2}}dj~{\cal A}_{(j,w)}^{{\bf s}_1}
{\cal A}_{(j^+,w^+)}^{{\bf s}_2}\chi_j^{+,w}(\tilde \theta,\tilde \tau,0)~
+~
ccr\, ,\nonumber
\eea
where $\Theta$ denotes the worldsheet CPT operator in the bulk theory, 
$\tilde q=e^{2\pi i\tilde 
\tau}$, $\tilde z=e^{2\pi i\tilde \theta}$, $\tilde\tau=-1/\tau$,
$\tilde\theta=\theta/\tau$, $(j^+,w^+)$ refer to the labels of the 
$(j,w)$-conjugate representations, $ccr$ denotes the contributions of
continuous representations
 and 
${\cal A}_{(j,w)}^{{\bf s}}$ 
are the Ishibashi coefficients of the boundary states.

The open string partition function for the ``spherical branes'' 
of the H$_3^+$  model was found in \cite{GKS} for $\theta=0$ and extended
to the case $\theta\neq0$ in \cite{RS}. It reads
\bea
Z_{{\bf s}_1{\bf s}_2}^{H_3^+}(\theta,\tau,0)=\sum_{J_3=\left|J_1-J_2\right|}^{J_1+J_2}~~
\chi_{J_3}(\theta,\tau,0)\, ,\label{Verldeg}
\eea
where ${\bf s}_i=\frac{\pi}{k-2}(1+2J_i)$ and $1+2J_i\in{\mathbb N}$. This
 reveals an open string spectrum of discrete degenerate representations.

The Lorentzian partition function is expected to reproduce 
that of the H$_3^+$ model after analytic continuation in $\theta$ and $\tau$.
Then, if we concentrate on the one-point functions of fields in discrete 
representations, we only need to consider the case 
$\theta+n\tau\not\in{\mathbb Z}$. 
Thus, using the generalized Verlinde formula 
(see appendix C for details), namely 
\bea
\sum_{J_3=\left|J_1-J_2\right|}^{J_1+J_2}~~
\chi_{J_3}(\theta,\tau,0)=\sum_{w=-\infty}^{\infty}
\int^{-\frac12}_{-\frac{k-1}{2}}
dj~~
\frac{S_{J_1}{}^{j,w}S_{J_2}{}^{j,w}}{S_{0}{}^{j,w}}~~
e^{2\pi i\frac k4 \frac{\theta^2}{\tau}}\chi_j^{+,w}(\frac\theta\tau,
-\frac1\tau,0)\, ,\label{verdeg}
\eea
we obtain the following expression for the coefficients of the boundary states:
\bea
{\cal A}_{(j,w)}^{{\bf s}}
=f(j,w)~(-)^w\sqrt{\frac{ 2}{ i}}\left(\frac{2}{k-2}
\right)^{\frac14}
\frac{\sin\left[{\bf s}\left(1+2j-w(k-2)\right)\right]}{\sqrt{\sin\left[\frac{\pi}{k-2}\left(1+2j\right)\right]}}\, ,
\label{A}
\eea
defined up to a function $f(j,w)$ satisfying 
$f(j,w)~f(-\frac k2-j,-w-1)=1$. 
\medskip

{\textbf{\textit{One-point functions}}}
\medskip

To find the one-point functions associated to these point-like
branes, let us make use 
of the following definition of boundary states (see for instance 
\cite{schomerus})\footnote{Strictly speaking,
 this identity is valid on a Euclidean worldsheet. However,
it is appropriate to use it here
since we want to explore the relation of our results
with those of the Euclidean model defined in 
\cite{israel} where the coefficients of the one-point functions are 
assumed to coincide with those of the Lorentzian AdS$_3$.}:
\bea
\left\langle \Phi^{(H)}\left(\left|j,m,\bar m,w\right\rangle;z,
\bar z\right)\right\rangle_{\bf s}=\left(\frac{d\xi}{dz}\right)^{\Delta_j} 
\left(\frac{d\bar\xi}{d\bar z}\right)^{\bar\Delta_j}
\left\langle 0\right|\Phi^{(P)}
\left(\left|j,m,\bar m,w\right\rangle;\xi,\bar \xi\right)
\left|{\bf s}\right\rangle\, ,\label{1pf}
\eea
where $\Phi^{(H)}\left(\left|j,m,\bar m,w\right\rangle;z,\bar z\right)$ 
($\Phi^{(P)}\left(\left|j,m,\bar m,w\right\rangle;\xi,\bar \xi\right)$) 
is the bulk field of the boundary (bulk) CFT
 corresponding to the state inside the 
brackets\footnote{Here $|j,m,\bar m,w>$ is a shorthand notation
 for $|j,m,w>\otimes|j,\bar m,w>$ and it must be distinguished from
 the orthonormal basis introduced in section \ref{coherentstates}.},
$z,\bar z$ denote the coordinates of the upper half plane and 
$\xi,\bar \xi$ those of the exterior of the unit disc. 

Conformal invariance forces the $l.h.s.$ of (\ref{1pf}) to be 
\bea
\left\langle \Phi^{(H)}\left(\left|j,m,\bar m,w\right\rangle;z,
\bar z\right)\right\rangle_{\bf s}=\frac{{\mathcal B}
({\bf s})_{m,\bar m}^{j,w}}
{|z-\bar z|^{\Delta_j+\bar\Delta_j}}\, ,
\eea
where the $z$-independent factor ${\cal B}({\bf s})_{m,\bar m}^{j,w}$ 
is not fixed by the conformal symmetry. The solution (\ref{Ishdef}), (\ref{Vdef}) implies
\bea
{\mathcal B}({\bf s})_{m,\bar m}^{j,w}=(-)^{j+m}\delta_{m,\bar m}
{\mathcal A}_{j,w}^{\bf s}~~,\label{resul}
\eea
from which the spectral flow symmetry determines $f=1$.

It is important to note that the normalization used here differs from 
the one usually considered in the literature. Our normalization is such that 
the spectral flow image of the primary operator
corresponding to the state 
 $|j,m,\bar m,w>$
is normalized to 1. 
In particular, it implies the following operator product expansions (OPE)
\bea
J^3(\zeta)\Phi^{(P)}\left(\left|j,m,\bar m,w\right\rangle;\xi,\bar \xi\right)
&=& 
\frac{m+\frac k2 w}{\zeta-\xi}\Phi^{(P)}\left(\left|j,m,\bar m,w\right\rangle;
\xi,\bar \xi\right)+\dots\cr\cr
J^{\pm}(\zeta)\Phi^{(P)}\left(\left|j,m,\bar m,w\right\rangle;\xi,\bar \xi
\right)&=&
 \frac{\sqrt{-j(1+j)+m(m\pm1)}}{\left(\zeta-\xi\right)^{1\pm w}}\Phi^{(P)}
\left(\left|j,m\pm1,\bar m,w\right\rangle;\xi,\bar \xi\right)\nonumber\\
&&+\dots~~~~\label{ope(p)}
\eea 
In appendix C, we  show that  (\ref{resul})
agrees with the one-point function obtained in \cite{israel}.

\subsubsection{{\it Cardy structure} in H$_2$ branes}

In appendix D we review the results for the one-point functions in
 maximally symmetric D-branes obtained by applying the method of \cite{israel}.
From the one-point functions
of fields in discrete 
representations  on H$_2$ branes we 
find the following Ishibashi coefficients (see (\ref{norm}) and (\ref{H2dis}))
\bea
{\cal A}_{(j,w)}^{\sigma'\equiv(j',w')} =
\frac{\pi}{\sqrt k}\left(\frac{2}{k-2}\right)^{\frac34}~ \frac{(-)^w e^{\frac{4\pi i}{k-2}(j'+\frac12-w'\frac{k-2}{2})(j+\frac12-w
\frac{k-2}{2})}}{\sqrt{\sin\left[\frac{\pi}{k-2}(2j+1)\right]}}\, ,
\eea 
satisfying  
\bea
{\cal A}_{(j,w)}^{j_1,w_1}~{\cal A}_{(j^+,w^+)}^{j_2,w_2}\sim 
(-)^{w_1+w_2}\frac{{\cal S}_{j_1w_1}{}^{jw}{\cal S}_{j_2w_2}{}^{j^+w^+}}
{{\cal S}_{0}{}^{jw}}\, ,
\eea 
where $\sim$ stands for equal up to the $k-$dependent factor 
$\frac{-i\, 4\,\pi^2}{k(k-2)}$. This expression leads
to the following degeneracy for the open string spectrum of discrete 
representations
\bea
{\cal N}_{j_1,w_1~j_2,w_2}{}^{j_3,w_3}=\frac{-i\, 2\,\pi^2(-)^{w_3}}{k(k-2)} \sum_{m=-\infty}^{\infty}\delta\left(j_2+j_3-j_1-\left(w_2+w_3-w_1\right)\frac{k-2}{2}+m\right)\, ,\nonumber
\eea
where the divergent integral $\int_0^1 d\lambda \frac{e^{-2\pi i 
\left(m+\frac12\right)\lambda }}{2i \sin(\pi \lambda)}$ has been 
replaced by its principal value, $\frac12$. 

Two comments are in order. First,
a non negative integer times a 
Kronecker or Dirac delta function would be expected for the degeneracy. 
An integer can be obtained through a small modification 
by an overall $k$-dependent factor in the one-point 
functions, but the sign factor $(-)^{w_3}$ cannot be removed 
in this way, and it inevitably  leads
to negative degeneracies. The second comment
 is about the Verlinde theorem. Contrary to what happens in RCFT, here
the {\it Cardy structure} is not accompanied by a Verlinde formula. 
Even, if we ignore the problems mentioned in the first comment, 
the naive application of this formula gives contributions 
to the fusion rules violating the spectral flow number 
conservation by an arbitrary amount, in contradiction with
 the selection rules determined
in \cite{mo3}.

\subsubsection{Coherent states for twined gluing conditions}

The gluing conditions defining the coherent states $|j,w\gg$
 for AdS$_2$ branes \cite{RR}, 
frequently called  twisted boundary conditions, are 
\bea
\left(J_n^3+\bar J_{-n}^3\right)\left|j,w\gg\right.=0, 
\qquad
\left(J_n^{\pm}+\bar J_{-n}^{\pm}\right)\left|j,w\gg\right.=0
\, .\label{bcads2}
\eea
These constraints are highly restrictive. As we show below, 
coherent states satisfying these conditions can only be found for  
representations where the holomorphic and  antiholomorphic sectors are 
conjugate of each other, $i.e.$ only for
 $w=0$, $\alpha=0,\frac12$  continuous representations   
in the AdS$_3$
model. 

Let us assume $|j,w\gg$ is an Ishibashi state 
 associated to the spectral flow image of a 
 discrete or  continuous 
representation. 
The spectral flow transformation (\ref{wtrans}) 
allows  to translate the problem of solving 
(\ref{bcads2}) 
to that of solving 
\bea
\left(J_n^3+\bar J_{-n}^3+kw\delta_{n,0}\right)\left|j\right\rangle^{w}=0\, 
,\qquad
\left(J_n^{\pm}+\bar J_{-n\mp2w}^{\pm}\right)\left|j\right\rangle^{w}=0
\, ,\label{gcads2}
\eea
where $\left|j\right\rangle^{w}=U_{-w}\bar U_{-w}|j,w\gg$ is 
in an unflowed representation\footnote{Notice that in the case $w=-\bar w$ 
discussed in \cite{RR} for the single covering of SL(2,$\mathbb R$),
one gets (\ref{bcads2}) with the unflowed $|j>^{w,\bar w}$ 
state replacing $|j,w\gg$ 
instead of (\ref{gcads2}).
 Then, once an Ishibashi state is found for $w=-\bar w=0$, 
the solutions for
 generic  representations with $w=-\bar w$ are trivially obtained applying the
spectral flow operation, and
coherent states in arbitrary spectral flow sectors are found.
This fails in AdS$_3$ and thus
 the discussion in $loc.~cit.$ does not apply here, except for
 $w=0$ discrete or $w=0$, $\alpha=0,
\frac12$ continuous representations.}.

The special case $n=0$  in (\ref{gcads2}) 
implies 
$2\alpha+kw~\in{\mathbb Z}$ and $ -2j+kw ~\in{\mathbb Z}$ 
for continuous and discrete representations, respectively.
In particular, for $w=0$ continuous representations there are two solutions 
with $\alpha=0,\frac12$, given by 
\bea
\left|j,0,\alpha\right.\gg= \sum_{n} \epsilon_n\bar U \left|j,w,\alpha;n,n\right\rangle, 
\eea
where the antilinear operator $U$ is defined by
\bea
U\prod_I J_{n_I}^{a_I}\left|j,m=\alpha,w=0\right\rangle=
\prod_I -J^{a_I}_{n_I}\left|j,m=-\alpha,w=0\right\rangle\, .
\eea 
It can be  easily verified that this defines an antiunitary operator and 
it is exactly the same Ishibashi state found in SU(2) \cite{Ishibashi}.

To understand why there are no solutions in other modules, let us expand 
the hypothetical Ishibashi state in the orthonormal base $|j,w,\zeta>\{|n>\otimes\overline{|m>}\}$, with $\zeta=\alpha$ or $+$ and $|n>,\overline{|m>}$ 
eigenvectors of $J_0^3,L_0$ and $\bar J_0^3,\bar L_0$ respectively. The constraint that Ishibashi states are annihilated by $L_0-\bar L_0$ forces  $|n>,\overline{|m>}$ to be at the same level.  
But taking into account that 
all modules at a given level are highest or lowest weight representations 
of the zero modes of the currents (with the only exception of $w=0$ 
continuous representations) and the fact that the eigenvalues of the highest 
(lowest) weight operators
 decrease (increase) after descending a finite number of levels, 
the first equation in (\ref{bcads2}) with $n=0$ 
has no solution
below certain level. This implies that below that level there 
are no contributions to the Ishibashi states and so, using for instance the 
constraint $(J_1^a+\bar J_{-1}^a)|j,w\gg=0$, 
it is easy to show by induction that 
no level contributes to the coherent states. 

The 
coherent states defined above are normalized as
\begin{equation}
\ll j,0,\alpha|e^{\pi i\tau\left(L_0+\bar L_0-\frac{c}{12}\right)}
e^{\pi i\theta(J_0^3-\bar J_0^3)}|j',0,\alpha'\gg ~= ~\delta (s-s')
\delta(\alpha-\alpha')~\chi_j^{\alpha,0}(\tau,\theta),\label{IshNormTwis}
\end{equation}
for $\alpha=0,\frac12$. The fact that it is only possible to construct
Ishibashi states associated to $w=0$ continuous representations 
is again in agreement with the one-point functions found in \cite{israel} 
and the conjecture in \cite{D} 
that only states in these representations couple to AdS$_2$ branes.

\section{Conclusions}

To conclude, let us summarize our results and
contrast them
with previous works in the literature.

We have computed the characters of the relevant representations of the AdS$_3$
model on the Lorentzian torus and studied their modular transformations.
We fully determined the generalized $S$ matrix, which
depends on the sign of $\tau$, and showed that
real modular parameters are crucial to
find the modular maps.

We have seen that the characters of continuous representations 
transform among themselves under  $S$  while 
both kinds of characters appear in the $S$ transformation of
 the characters of
discrete representations. An important
 consequence of this fact is that the Lorentzian partition function is not 
 modular invariant 
(and the departure from modular invariance is not just the sign appearing in
(\ref{ZDmod})). 
The analytic continuation to obtain the 
Euclidean partition function (which must be invariant) is 
not fully satisfactory. Following the road of \cite{mo1} and simply 
discarding the contact terms, one recovers the partition function of the
H$_3^+$ model obtained in \cite{gawe}. But even 
though modular invariant, this expression has poor information about the 
spectrum. 
 Starting from the partition function of the SL(2,$\mathbb R$)/U(1)
coset computed in 
\cite{hpt} and using path integral techniques,
an alternative expression 
was found in \cite{kounnas}. Although formally
divergent, it is modular invariant and allows to 
read  the spectrum of the model\footnote{The spectrum was also 
obtained from a computation
 of the Free Energy in \cite{mo2}.}. 
It was shown that the partition function
obtained in \cite{gawe,mo1} is recovered after some 
formal manipulations.
It would be interesting to better understand how the information is lost
in the procedure implemented in \cite{kounnas}
 and to explore if it is possible to find an analytic continuation of the 
Lorentzian partition function
leading to the integral expression obtained in 
$loc. cit.$ (or an equivalent one), in a  controlled 
way in which the knowledge  on the spectrum is not 
lost.

The treatment of the boundary states presented in section 4 differs from
 previous works.
While we have expressed them
 as a sum over Ishibashi states, in  other related 
 models
such as 
H$_3^+$ \cite{PST}, Liouville \cite{FZZ} or 
the Euclidean 
black hole \cite{RS}, the boundary states have been expanded, instead,
in terms of
primary states and their descendants. The coefficients in the
latter expansions 
directly give the one-point functions of the primary fields. 
For instance, 
in the H$_3^+$ model, the gluing conditions were imposed in \cite{PST}
not over the 
Ishibashi states but 
over the one-point functions. 
One of the reasons why this approach seems more 
suitable for H$_3^+$ is the observation that the 
expectation values used to fix the normalization of the Ishibashi states 
diverge in the hyperbolic model\footnote{Notice that, 
contrary to the AdS$_3$ or SU(2) models, the 
continuous representations appearing in the Hilbert space of the H$_3^+$ 
 model do not factorize as  tensor products of a holomorphic times an 
antiholomorphic representation. So, instead of the characters of the 
holomorphic sector appearing for instance in (\ref{IshNorm}), the analog ones 
in 
the hyperbolic model have a trace over certain subspace of states satisfying 
$J_0^3=\pm \bar J_0^3$, depending on the gluing conditions considered. And 
this trace is divergent.}.    
As we have seen, this is not the case in AdS$_3$.

The generalization of the Verlinde formula proposed in section 4
gives 
the fusion rules
 of the degenerate 
representations of SL(2,${\mathbb R}$) appearing in the spectrum of open
strings attached to the point-like D-branes of the model and
the coefficients of their 
boundary states. The formula holds for generic $\theta, \tau$  far from
$\theta +n\tau\in{\mathbb Z}$. It would be interesting  to study the extension
to generic $\theta,\tau$ which requires to consider the $S$ matrix block
(\ref{degs}). Furthermore, one could also study the modular transformations
of the characters of other degenerate representations and their spectral flow
images and
explore the validity of generalized Verlinde formulas  in these cases.

We have shown that
the one-point functions of fields in discrete representations coupled to
H$_2$ branes are determined by one of the diagonal blocks of
the generalized $S$ matrix, as usual in RCFT. However, 
a puzzle arises when considering
the open/closed duality which gives
negative degeneracies in the open string spectrum of these branes.
In constrast to general expectations, here the {\it Cardy structure} is not
 accompanied by a Verlinde theorem. Moreover,
 the Verlinde-like formula does not give the
fusion rules of  the bulk AdS$_3$ 
model. In particular,
besides some undesirable negative signs, it gives
 contributions of arbitrary spectral flow numbers to the fusion 
of states in discrete representations, thus violating the selection rules
established in \cite{mo3}.
Much remains to be understood on the role of the Verlinde theorem (or
suitable generalizations) in non RCFT. In particular,
more work is necessary to put the fusion rules of the AdS$_3$ WZNW model
on a firmer ground, as the mechanism determining the truncation of states in
the operator algebra is far from elucidated.

\subsection*{Acknowledgments} We would like to thank
Carlos Cardona, Horacio Falomir,
Sergio Iguri, Juan Maldacena, Jorge Russo,  Yuji Satoh and especially
Silvain Ribault
 and Jan Troost for valuable
discussions.
This work was supported by grants 
PIP CONICET  112 200801 00507 and
UBACyT X161.

\addcontentsline{toc}{section}{Appendices}

\section*{Appendices}

\appendix
\section{The Lorentzian torus}

In this appendix we present a description of
 the moduli space of 
the torus with  Lorentzian  metric\footnote{Tori in $1+1$ dimensions have 
been considered previously in 
\cite{moore} - \cite{russo} in the context of string propagation 
in time dependent 
backgrounds.}. Although it can be easily obtained from the 
Euclidean case, we include it here for completeness.

Consider the two dimensional torus with 
worldsheet coordinates $\sigma^1, \sigma^2$ 
obeying the identifications
\begin{equation}
(\sigma^1,\sigma^2)\cong(\sigma^1+2\pi n,\sigma^2+2\pi m),~~~~~~n,m
\in{\mathbb Z}. \label{pertoro}
\end{equation}
By diffeomorphisms and Weyl transformations that leave invariant the 
periodicity, 
a general two dimensional Lorentzian metric can be taken to the form
\begin{equation}
ds^2=(d\sigma^1+\tau_+d\sigma^2)(d\sigma^1+\tau_-d\sigma^2),\label{gToroLor}
\end{equation}
where
$\tau_+,\tau_-$ are two real independent parameters.
Recall that the metric of the Euclidean torus, 
namely $ds^2=|d\sigma^1+\tau d\sigma^2|^2$, is
degenerate for $\tau\in\mathbb R$ since $det~g=(\tau-\tau^*)^2$. 
In contrast, here it is degenerate for $\tau_-=\tau_+$.

The linear transformation
\begin{eqnarray}
\tilde\sigma^1=\sigma^1+\tau^+\sigma^2,\quad
\tilde\sigma^2=\tau^-\sigma^2\, , \qquad \tau^{\pm}=\frac{\tau_-\pm\tau_+}{2},
\end{eqnarray}
takes (\ref{gToroLor}) to the Minkowski metric.
The new coordinates obey the periodicity conditions
\begin{equation}
(\tilde\sigma^1,\tilde\sigma^2)
\cong(\tilde\sigma^1+2\pi n+2\pi m\tau^+,\tilde\sigma^2+2\pi 
\tau^-m)\, ,~~~~~~n,m\in{\mathbb Z}\, ,\label{pertoroMink}
\end{equation}
while the light-cone coordinates
$\tilde \sigma_{\pm}=\tilde\sigma^1\pm\tilde\sigma^2$, obey
\begin{equation}
\tilde \sigma_{\pm}\cong\tilde \sigma_{\pm}+2\pi n+2\pi m\tau_{\mp}\, .
\end{equation}

 In the Euclidean case, 
there are in addition
 global transformations that cannot be smoothly connected to the 
identity, generated by Dehn twists. 
A twist along the $a$ cycle of a Lorentzian torus
preserves the metric (\ref{gToroLor}) but
changes the periodicity  to
\begin{equation}
(\tilde\sigma^1,\tilde\sigma^2)\cong(\tilde\sigma^1+2\pi n+2
\pi m(1+\tau^+),\tilde\sigma^2+2\pi m~\tau^-)\, ,~~~~~~n,m\in{\mathbb Z},
\end{equation}
or
\begin{equation}
\tilde \sigma_{\pm}
\cong \tilde \sigma_{\pm}+2\pi n+2\pi m~(\tau_{\mp}+1)\, .
\end{equation}
Thus it gives a torus with modular parameters
 $(\tau_+',\tau_-')=(\tau_++1,\tau_-+1)$.
A twist along the $b$ cycle leads to the following periodicity conditions
\begin{equation}
(\tilde\sigma^1,\tilde\sigma^2)\cong(\tilde\sigma^1+2\pi n(1+\tau^+)+2\pi 
m~\tau^+,\tilde\sigma^2+2\pi n~\tau^-+2\pi m~\tau^-),~~~~~~n,m\in{\mathbb Z},
\end{equation}
or
\begin{equation}
\tilde \sigma_{\pm}\cong\tilde \sigma_{\pm}+2\pi 
n(1+\tau_{\mp})+2\pi m~\tau_{\mp}\, .
\end{equation}
As in the Euclidean case, this is equivalent to a torus with 
$(\tau_+',\tau_-')=(\frac{\tau_+}{\tau_++1},\frac{\tau_-}{\tau_-+1})$ 
and conformally flat metric. But there is a crucial difference. In the 
Euclidean case, the overall conformal factor multiplying the flat metric 
is positive definite, namely $\frac{1}{(1+\tau)(1+\tau^*)}$. 
On the contrary, in the 
Lorentzian torus, the conformal factor $\frac{1}{(1+\tau_-)(1+\tau_+)}$ is not 
positive definite and so, it can not be generically eliminated through a
Weyl transformation.

Defining the modular $S$ transformation as 
$S\tau_{\pm}=-\frac{1}{\tau_{\pm}}$, we can write
$\tau'_{\pm}=\frac{\tau_\pm}{1+\tau_\pm}=TST~ \tau_\pm$, and then
the problem can be reformulated in the following way. 
The  $T$ transformation works as in the Euclidean case. 
Instead, under a modular $S$ transformation, the torus 
defined by (\ref{pertoro}) and (\ref{gToroLor}) 
is equivalent to a torus with the same 
periodicities but with the following metric (after diffeomorphisms and
Weyl rescaling)   
\bea
ds^2=sgn(\tau_-\tau_+)\left(d\sigma '^1+\tau_+d\sigma'^2\right)
\left(d\sigma '^1+\tau_-d\sigma '^2\right)\, .
\eea

\medskip

{\bf \itshape\large The fundamental region}

\medskip
In the Euclidean torus,
one can find a coordinate system preserving
the periodicity conditions (\ref{pertoro}),  where the metric takes the form
$ds^2=|d\sigma^1+\tau d\sigma^2|^2$, with $\tau\in {\mathbb C}$. Since it is
 invariant under complex conjugation,
 the complex
$\tau$ plane can be restricted to
Im $\tau>0$ (discarding Im$ \tau=0$ because it gives a degenerate metric).
Similarly, in the Lorentzian case, the metric (\ref{gToroLor})
is invariant under
$\tau_+\leftrightarrow\tau_-$ and one can take
$\tau_+>\tau_-$ 
(discarding $\tau_+=\tau_-$).

Unlike the Euclidean case, where the $S$ transformation maps the interior
to the exterior of the unit circle, 
in the Lorentzian case it maps
the interior of the
hyperbola $\tau_+=-\tau_-^{-1}$ in the second
quadrant to the exterior of the hyperbola in the fourth quadrant.
But the symmetry
$\tau_+
\leftrightarrow\tau_-$, allows to identify this  region of the fourth quadrant
 with the exterior of the
hyperbola in the second quadrant. Similarly, using this symmetry, the
$S$ transformation maps the exterior 
to the interior of the hyperbola in the second quadrant
 (see the figure) and leaves
the points
on the hyperbola 
fixed. One of these points is
 $(\tau_-,\tau_+)=(-1,1)$ which corresponds to the Minkowski metric.
 (Recall that in the Euclidean case there is a single fixed point,
$\tau=i$, giving a flat Euclidean metric).


\centerline{\psfig{figure=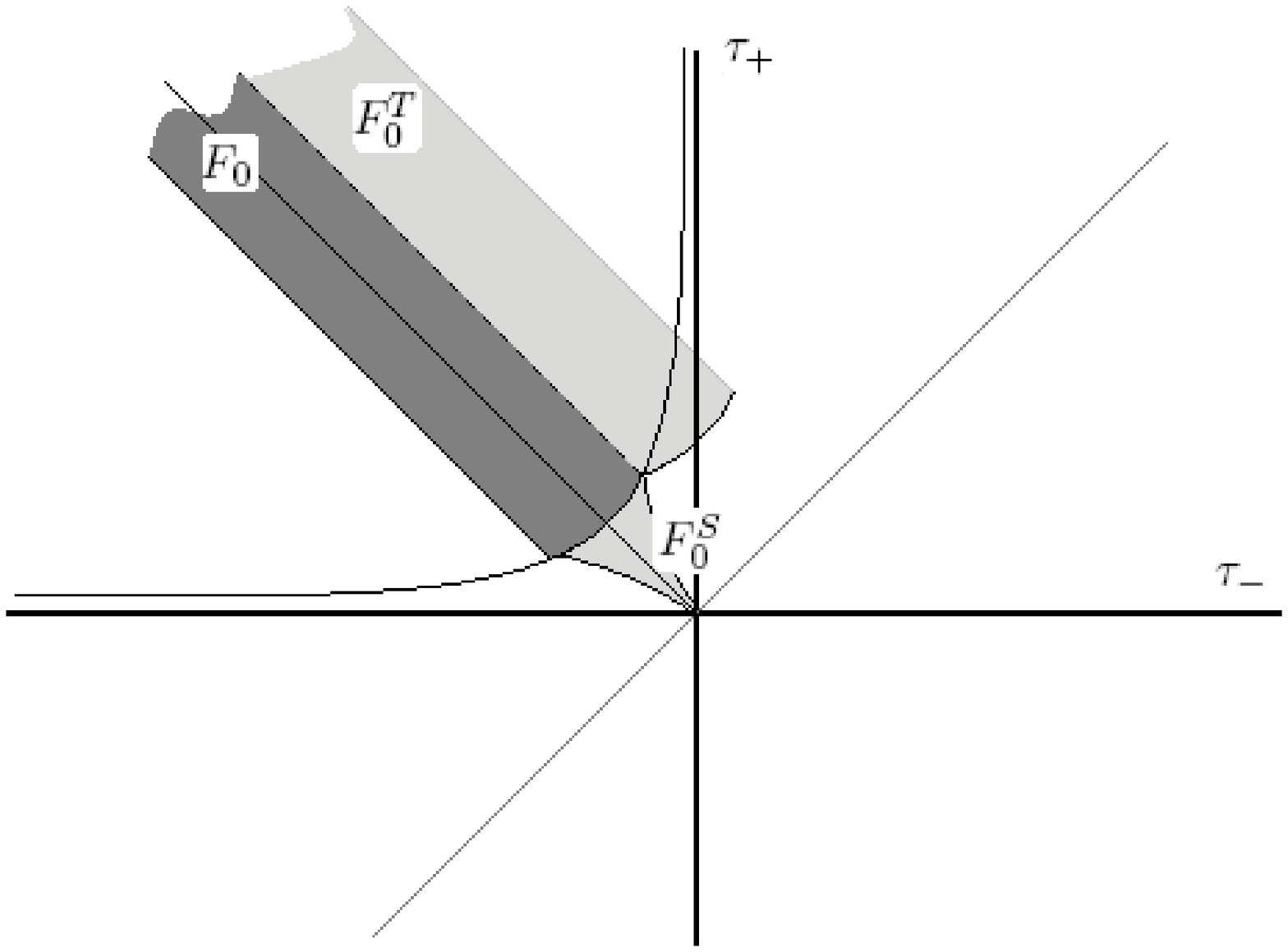,width=10cm}}
 {\footnotesize
{Figure 2: A
fundamental region $F_0$ can be defined as
$-\tau_- -1\leq\tau_+<-\tau_- +1$, $\tau_-<0,~\tau_+\geq-
\tau_-^{-1}$ ($\tau_+>-\tau_-^{-1}$) for $\tau_+>-\tau_-$ ($\tau_+<-\tau_-$). 
Other possible 
fundamental regions are  the images of $F_0$ by $S$ or $T$, 
denoted $F_0^S$, $F_0^T$ respectively.}}


\section{The mixing block of the $S$ matrix}

In this appendix we sketch the computation of the off-diagonal
 block of the $S$ matrix mixing the characters of 
continuous and discrete representations.

\medskip
{\textbf{\textit{ A useful identity}}}\label{MixBlockApp}
\medskip
 
It is convenient to begin displaying a useful identity.

Let 
$h(x;\epsilon_0)=\frac{1}{1-e^{2\pi i (x+i\epsilon_0)}}$, with
$x\in {\mathbb R}$,  be 
the distribution defined as the weak limit $\epsilon_0\rightarrow0$ 
and  $G(x;\epsilon_1,\epsilon_2,\epsilon_3,\dots)$ a generalized function 
having simple poles outside of the real line\footnote{
$G(x;0,0,0,\dots)$ not necessarily has only simple poles. 
In the most general case, it will have poles of arbitrary order.},
defined as the weak limit $\epsilon_i\rightarrow0, i=1,2,3,...$.  
The non vanishing infinitesimals 
 $\epsilon_i$ are allowed to depend on the $x$ coordinate
and they all differ from each other
in an open set around each simple pole. Then, the following identity holds 
(in a distributional sense):
\bea
&&\frac{1}{1-e^{2\pi i (x+i\epsilon_0)}} G(x;\epsilon_1,\epsilon_2,
\epsilon_3,\dots)~~=~~
\frac{1}{1-e^{2\pi i (x+i\tilde\epsilon_0)}} G(x;\epsilon_1,\epsilon_2,
\epsilon_3,\dots)\cr\cr
&&\ \ \ \ \ \ \ \ \ \ \ \ \ \ \ \ \ \ \ \ \ \ \ 
+~~\sum_{x_i^\downarrow}\delta(x-x_i^\downarrow) G(x;\epsilon_1-\epsilon_0,
\epsilon_2-\epsilon_0,\epsilon_3-\epsilon_0,\dots)\cr
&& \ \ \ \ \ \ \ \ \ \ \ \ \ \ \ \ \ \ \ \ \ \ \ 
-~~\sum_{x_i^\uparrow}\delta(x-x_i^\uparrow) G(x;\epsilon_1-\epsilon_0,
\epsilon_2-\epsilon_0,\epsilon_3-\epsilon_0,\dots),\label{identdeltas}
\eea
where $\tilde\epsilon_0$ is a new infinitesimal parameter, 
$x_i^\downarrow$ ($x_i^\uparrow$) is the real part of the pulled down (up)
poles, $i.e.$ those poles where 
$\epsilon_0(x_i^\downarrow)<0<\tilde\epsilon_0(x_i^\downarrow)$ 
($\tilde\epsilon_0(x_i^\uparrow)<0<\epsilon_0(x_i^\uparrow)$). 
Of course, here $x_i^\downarrow,\,x_i^\uparrow\in{\mathbb Z}$, 
but 
(\ref{identdeltas}) can be trivially generalized to other 
functionals 
 having simple poles, the only change being that the residue
has to multiply each 
delta function. 

The proof of this identity 
follows from multiplying (\ref{identdeltas}) by an arbitrary test 
function ($f(x)\in C_0^{\infty}$) and integrating  over the real line.

As an example, let us consider the simplest case $G=1$, $\epsilon_0=0^+$, 
$\tilde\epsilon_0=0^-$, where one recovers the well known formula
\bea
\frac{1}{1-e^{2\pi i(x+i0^+)}}=\frac{1}{1-e^{2\pi i(x+i0^-)}}
-\sum_{m=-\infty}^{\infty}\delta(x+m)\,.
\eea

{\textbf{\textit{The mixing   
block}}}
\medskip

Let us first consider the modular transformation of the
 elliptic theta function
\bea
\frac{1}{i\vartheta_{11} (\theta+i\epsilon_2^w,\tau+i\epsilon_1)}
&\rightarrow &
\frac{1}{i\vartheta_{11} (\frac\theta\tau+i\epsilon_2^w,
-\frac1\tau+i\epsilon_1)}\equiv
\frac{1}{i\vartheta_{11} (\frac{\theta+i\epsilon'_2{}^w}{\tau+i\epsilon'_1},
-\frac{1}{\tau+i\epsilon'_1})}\cr
&=&\frac{-sgn(\tau)e^{-\pi i \frac{\theta^2}{\tau}}e^{-sgn(\tau)i\frac\pi4}}
{\sqrt{|\tau|}}
\frac{1}{\vartheta_{11} (\theta+i\epsilon'_2{}^w,\tau+i\epsilon'_1)}, 
\label{thetatrans}\\
&&\left\{\begin{array}{lcr}
~\epsilon'_1=\tau^2\epsilon_1~,\cr
\epsilon'_2{}^w=\tau\left(\epsilon_2^w+\theta\epsilon_1\right)~,
\end{array}\right.
\eea
and $\epsilon_1,\epsilon_2^w$ satisfy (\ref{epsiloncond}).
The identity (\ref{stheta}) was used in the last line of (\ref{thetatrans})
and the limits $\epsilon'_1,\epsilon'_2{}^w\rightarrow0$ 
were taken where  it is allowed.  

Let us now concentrate on the last term in (\ref{thetatrans}). 
It is explicitly given by  (\ref{theta11b}), 
where now the $\epsilon$'s are replaced by $\epsilon'_1,\epsilon'_3{}^{n,w},
\epsilon'_4{}^{n,w}$ satisfying $\epsilon'_1>0$,
\bea
\epsilon'_3{}^{n,w}\left\{\begin{array}{lcr}
<0\,~,\theta-n\tau\leq-1-w\cr
>0\,,~\theta-n\tau\geq-w
\end{array}\right.,~~~~~
\epsilon'_4{}^{n,w}\left\{\begin{array}{lcr}
<0\,~,\theta+n\tau\geq-w\cr
>0\,,~\theta+n\tau\leq-1-w
\end{array}\right. \, ,\ \ \ \tau<0 ,\\
\epsilon'_3{}^{n,w}\left\{\begin{array}{lcr}
<0\,~,\theta-n\tau\geq-w\cr
>0\,,~\theta-n\tau\leq-1-w
\end{array}\right. ,~~~~~
\epsilon'_4{}^{n,w}\left\{\begin{array}{lcr}
<0\,~,\theta+n\tau\geq-w\cr
>0\,,~\theta+n\tau\leq-1-w
\end{array}\right.\, , \ \ \ \ \tau>0\, .
\eea

By comparing with  (\ref{epsilon34})
and using (\ref{identdeltas}), one finds, for instance in the case $w<0,
\tau<0$, after a straightforward but tedious computation, the following 
identity:
\bea
&&\frac{1}{i\vartheta_{11} (\theta + i\epsilon '_2{}^w, \tau +i \epsilon '_1)}
~=~
\frac{1}{i\vartheta_{11} (\theta + i\epsilon_2{}^w, \tau +i \epsilon_1)}\cr
&&\ \ \ \ \
-\frac{1}{\eta^3(\tau+i\epsilon_1)}\left[e^{-i\pi\theta}\sum_{n=0}^{\infty}
\sum_{m=-\infty}^{-w-1}(-)^n e^{\pi i\tau n(1+n)}\delta(-\theta+n\tau+m)
\right. \nonumber\\
&& \ \ \ \ \ \ \ \left.\qquad \qquad~~
+~e^{i\pi\theta}\left(\sum_{n=1}^{-w-1}\sum_{m=w+1}^{\infty}-
\sum_{n=-w}^{\infty}\sum_{m=-\infty}^{w}\right)(-)^n e^{\pi i\tau n(1+n)}
\delta(\theta+n\tau+m)\right]\, .\nonumber
\eea
Repeating the same analysis for the other cases one finds, for arbitrary 
$w$,
\bea
\frac{1}{i\vartheta_{11} (\theta+ i\epsilon'_2{}^w, \tau+i \epsilon'_1)}
&=&\frac{1}{i\vartheta_{11} (\theta+ i\epsilon_2{}^w, \tau+i \epsilon_1)}\cr
&+&
\left[~\sum_{n=-\infty}^{w}\left\{\begin{array}{lcr}
\sum_{m=-\infty}^{w}\delta(\theta-n\tau+m),\tau<0\cr
\sum_{m=1+w}^{\infty}\delta(\theta-n\tau+m),\tau>0
\end{array}\right. \right.\cr
&&-
\left.
\sum_{n=1+w}^{\infty}\left\{\begin{array}{lcr}
\sum_{m=1+w}^{\infty}\delta(\theta-n\tau+m),\tau<0\cr
\sum_{m=-\infty}^{w}\delta(\theta-n\tau+m),\tau>0
\end{array}\right.\right]\frac{(-)^{n+m}e^{2i\pi \tau \frac{n^2}{2}}}
{\eta^3(\tau+i\epsilon_1)}\nonumber
\eea

Using (\ref{gaussian}) and summing or subtracting delta function terms 
like in (\ref{chiw<w'}) and (\ref{chiw>w'}), 
in order to construct the characters of discrete representations,
one finds 
\bea
&&\chi_j^{+,w}(\frac\theta\tau,-\frac1\tau,0)~=~
e^{-2\pi i\frac k4\frac{\theta^2}{\tau}}sgn(\tau)\cr\cr
&&~~~~~~\times~\left\{\sum_{w'=-\infty}^{\infty}
\int^{-\frac{1}{2}}_{-\frac{k-1}{2}}dj'
\sqrt{\frac{2}{k-2}}(-)^{w+w'+1} e^{\frac{4\pi i}{k-2} 
\left(j'+\frac12-w'\frac{k-2}{2}\right)\left(j+\frac12-
w\frac{k-2}{2}\right)} \chi_{j'}^{+,w'}(\theta,\tau,0)\right.\cr\cr
&&~~~~~~
+\sum_{w',n,m~\in~ {\cal I(\tau)}}~ \int^{-\frac{1}{2}}_{-\frac{k-1}{2}}dj'
\sqrt{\frac{2}{k-2}}(-)^{w+1} e^{\frac{4\pi i}{k-2} \left(j'+\frac12-w'\frac{k-2}{2}\right)\left(j+\frac12-
w\frac{k-2}{2}\right)}  \cr\cr
&&~~~~~~~~~~~~~~
\times ~\left.\frac{e^{-\frac{2\pi i}{k-2}\tau(j'+\frac12-w'
\frac{k-2}{2})^2} 
e^{-2\pi i\theta(j'+\frac12-w'\frac{k-2}{2})}}{\eta^3(\tau+i\epsilon_1)} 
(-)^{n+m} e^{2\pi i\tau \frac{n^2}{2}}
\delta(\theta-n\tau+m)\right\}\, ,\nonumber
\eea
where $\sum_{w',n,m~\in~ {\cal I(\tau)}}$ is expected to reproduce 
the contribution from the 
continuous representations and is explicitly given by
\bea
\sum_{w',n,m~\in~ {\cal I(\tau)}} &\equiv&-\sum_{w'=-\infty}^{w-1}
\sum_{n=1+w'}^{w}\sum_{m=-\infty}^{\infty}+\sum_{w'=1+w}^{\infty}
\sum_{n=1+w}^{w'}\sum_{m=-\infty}^{\infty}\cr\cr
&&+~\sum_{w'=-\infty}^{\infty}\left(
\sum_{n=-\infty}^{w}\left\{\begin{array}{lcr}
\sum_{m=-\infty}^w\cr
\sum_{m=1+w}^\infty
\end{array}\right. -
\sum_{n=1+w}^{\infty}\left\{\begin{array}{lcr}
\sum_{m=1+w}^\infty\cr
\sum_{m=-\infty}^w
\end{array}\right.\right)\cr\cr
&=&\sum_{w'=-\infty}^\infty\left(
\sum_{n=-\infty}^{w'}\left\{\begin{array}{lcr}
\sum_{m=-\infty}^w\cr
\sum_{m=1+w}^\infty
\end{array}\right. -
\sum_{n=1+w'}^{\infty}\left\{\begin{array}{lcr}
\sum_{m=1+w}^\infty\cr
\sum_{m=-\infty}^w
\end{array}\right.\right)\cr\cr
&=&\sum_{n=-\infty}^\infty~\left(
\sum_{w'=n}^{\infty}~\left\{\begin{array}{lcr}
\sum_{m=-\infty}^w\cr
\sum_{m=1+w}^\infty
\end{array}\right. -
\sum_{w'=-\infty}^{n-1}\left\{\begin{array}{lcr}
\sum_{m=1+w}^\infty\cr
\sum_{m=-\infty}^w
\end{array}\right.\right)\, ,\nonumber
\eea
where the upper lines inside the brackets hold
 for $\tau <0$ and the lower ones
for $\tau >0$.
In the last line we have exchanged the order of summations. 
The sum over $w'$ together with the integral over $j'$, the spin of the states
 in discrete representations, match together 
to give, after analytic continuation, the integral over $s'$,
the imaginary part of the spin of the states in the principal continuous 
representations: 
\newpage
\bea
&&\sum_{w'=n}^{\infty} \int^{-\frac{1}{2}}_{-\frac{k-1}{2}}dj'
e^{\frac{4\pi i}{k-2} \left(j'+\frac12-w'\frac{k-2}{2}\right)\left(j+\frac12-
w\frac{k-2}{2}\right)} e^{-\frac{2\pi i}{k-2}\tau(j'+\frac12-w'
\frac{k-2}{2})^2} 
e^{-2\pi i\theta(j'+\frac12-w'\frac{k-2}{2})}\cr
&&~~~=~\int_{-\infty}^{0}d\lambda
e^{\frac{4\pi i}{k-2} \left(\lambda-n\frac{k-2}{2}\right)\left(j+\frac12-
w\frac{k-2}{2}\right)} e^{-\frac{2\pi i}{k-2}\tau(\lambda-n\frac{k-2}{2})^2} 
e^{-2\pi i\theta(\lambda-n\frac{k-2}{2})}\cr\cr
&&~~~=\left\{\begin{array}{lcr}
~i\int_0^\infty ds' ~e^{\frac{4\pi i}{k-2} \left(-is'-n\frac{k-2}{2}\right)
\left(j+\frac12-
w\frac{k-2}{2}\right)} e^{-\frac{2\pi i}{k-2}\tau(-is'-n\frac{k-2}{2})^2} 
e^{-2\pi i\theta(-is'-n\frac{k-2}{2})} ,~\tau<0~,\cr\cr
-i\int_0^\infty ds' ~e^{\frac{4\pi i}{k-2} \left(is'-n\frac{k-2}{2}\right)
\left(j+\frac12-
w\frac{k-2}{2}\right)}~ e^{-\frac{2\pi i}{k-2}\tau(is'-n\frac{k-2}{2})^2} 
~e^{-2\pi i\theta(is'-n\frac{k-2}{2})}~ ,~\tau>0~.
\end{array}\right.\nonumber
\eea 
After a similar analysis for the terms 
in the sum $\sum_{w'=-\infty}^{n-1}$ and 
relabeling the dummy index $n\rightarrow w'$, one finds the 
following contribution from the continuous series    
\bea
\sum_{w'=-\infty}^{\infty}~~ i\sqrt{\frac{2}{k-2}}\int_{0}^{\infty} ds'
(-)^{w+w'+1} \left[\sum_{m=-\infty}^{w}e^{\frac{4\pi i}{k-2} \left(-is'-w'
\frac{k-2}{2}\right)\left(j+\frac12-
w\frac{k-2}{2}\right)}e^{-2\pi i m\left(\frac12+is'+w'\frac{k-2}{2}\right)}
\right.  \cr\cr
-\left. \sum_{m=1+w}^{\infty}e^{\frac{4\pi i}{k-2} \left(is'-w'\frac{k-2}{2}
\right)\left(j+\frac12-
w\frac{k-2}{2}\right)}e^{-2\pi i m\left(\frac12-is'+w'\frac{k-2}{2}\right)}
\right]               
\frac{e^{2\pi i\tau\left(\frac{s'^{2}}{k-2}+\frac k4 w'{}^2\right)}}
{\eta^3(\tau+i\epsilon_1)}\delta(\theta-w'\tau+m)\nonumber
\eea

Finally, using (\ref{conttrick}), with the appropriate relabeling and 
performing the sum over $m$ (which then simply  reduces 
to a geometric series)
one gets
\bea
\sum_{w'=-\infty}^{\infty}\int_0^\infty ds'\int_0^1d\alpha' 
{\cal S}_{j,w}{}^{s',\alpha',w'} \chi_{s'}^{\alpha',w'}(\theta,\tau,0),
\eea
with
\bea
{\cal S}_{j,w}{}^{s',\alpha',w'}=-i\sqrt{\frac{2}{k-2}}e^{-2\pi i\left(w'j-w
\alpha'-ww'\frac k2\right)} \left[\frac{e^{\frac{4\pi}{k-2}s'\left
(j+\frac12\right)}}{1+e^{-2\pi i(\alpha'-is')}} + \frac{e^{-\frac{4\pi}{k-2}s'
\left(j+\frac12\right)}}{1+e^{-2\pi i(\alpha'+is')}}\right]\, .~~~~
\eea
\medskip

It is interesting to note that (repeated indices denote implicit sum)
\bea
{\cal S}_{j,w}{}^{s_1,\alpha_1,w_1} ~~
{\cal S}_{s_1,\alpha_1,w_1}{}^{s',\alpha',w'}&=&
-~{\cal S}_{j,w}{}^{j_1,w_1}~~{\cal S}_{j_1,w_1}{}^{s',\alpha',w'}\cr
&=&\frac{(-)^{w+w'+1}}{2\pi}\sum_{m=-\infty}^{\infty}\left[\frac{1}
{\frac12+\alpha '-is'-m}+ \frac{1}{\frac12+\alpha'+is'-m} \right]\cr 
&&\times ~ \delta\left(j-\alpha'-(w+w')\frac{k-2}{2}+m\right)\, .
\label{S2}
\eea

The first line implies ${\cal S}^2_{~j,w}{}^{s',\alpha',w'}=0$.

To show that $({\cal S}T)^3_{~j,w}{}^{s',\alpha',w'}=0$ is a bit more involved. This block is explicitly given by
\bea
{\cal S}_{j,w}{}^{s_1,\alpha_1,w_1}\left[  \left(T{\cal S} T{\cal S} T\right)_{s_1,\alpha_1,w_1} {}^{s',\alpha',w'}\right] +
 {\cal S}_{j,w}{}^{j_1,w_1}\left[ \left(T{\cal S} T {\cal S} T\right)_{j_1,w_1}{}^{s',\alpha',w'} \right]\, .
\eea
The first term above coincides with the first one in (\ref{S2}). This is a consequence of (\ref{STcontprop}), which implies $ \left(T{\cal S} T{\cal S} T\right)_{s_1,\alpha_1,w_1} {}^{s',\alpha',w'}= {\cal S}_{s_1,\alpha_1,w_1} {}^{s',\alpha',w'}$. So, in order for this block to vanish it is sufficient to show that the term inside the second bracket is exactly the ${\cal S}$ matrix mixing block. 

The factor inside the last bracket splits into the sum
\bea
&&T_{j_1,w_1}{}^{j_2,w_2}{\cal S}_{j_2,w_2}{}^{s_3,\alpha_3,w_3} T_{s_3,\alpha_3,w_3}{}^{s_4,\alpha_4,w_4} {\cal S}_{s_4,\alpha_4,w_4}{}^{s_5,\alpha_5,w_5} T_{s_5,\alpha_5,w_5}{}^{s',\alpha',w'} \cr 
&&~+ ~~~~ T_{j_1,w_1}{}^{j_2,w_2}{\cal S}_{j_2,w_2}{}^{j_3,w_3} T_{j_3,w_3}{}^{j_4,w_4} {\cal S}_{j_4,w_4} {}^{s_5,\alpha_5,w_5} T_{s_5,\alpha_5,w_5}{}^{s',\alpha',w'}\, .\label{ST3d-c}
\eea
These terms are very difficult to compute separately because each 
one gives the integral of a Gauss error function. So, we show 
here how  the sums  can be reorganized
in order to cancel all the intricate 
integrals when summing both terms and one ends with the mixing block 
${\mathcal S}_{j_1,w_1}{}^{s',\alpha',w'}$. In fact, after some few steps, 
the first line can be expressed as 
\bea
&&\sqrt{\frac{2}{k-2}}\int_0^\infty ds~ \left\{ \tilde S_{j_1,w_1}{}^{s_2,\alpha_2,w_2} \left[\sum_{w=-\infty}^0 e^{-i\frac\pi4}e^{-\frac{2\pi i}{k-2}\left[-is-w\frac{k-2}{2}-(j_1+\frac12)+is_2\right]^2} e^{2\pi i w\left(\alpha_2+\frac12-is_2\right)}\right.\right.\cr
&&~~~~~~-~~~~\left.\left. \sum_{w=1}^\infty e^{-i\frac\pi4}e^{\frac{2\pi i}{k-2}\left[-is-w\frac{k-2}{2}-(j_1+\frac12)+is_2\right]^2} e^{2\pi i w\left(\alpha_2+\frac12-is_2\right)}\right] + (s_2\rightarrow -s_2) \right\},~~\label{ST3a}
\eea
where we have introduced $~~\displaystyle \tilde S_{j_1,w_1}{}^{s_2,\alpha_2,w_2}=-i\sqrt{\frac{2}{k-2}}~ e^{-2\pi i\,\left(w_2j_1-w_1\alpha_2-w_1w_2\frac k2\right)}~ e^{\frac{4\pi s_2}{k-2}\left(j_1+\frac 12\right)}$.

On the other hand, the second line in (\ref{ST3d-c}) takes the form
\bea
&&\sum_{w=-\infty}^{\infty}\sqrt{\frac{2}{k-2}}\int_{-\frac{k-1}{2}}^{-\frac12} dj~e^{i\frac\pi2} e^{-\frac{2\pi i}{k-2}\left[j+\frac12-w\frac{k-2}{2}-(j_1+\frac12)+is_2\right]^2} e^{2\pi i w\left(\alpha_2+\frac12-is_2\right)} \cr &&~~~~~~~~~~~~~\times~~~~ \frac{\tilde S_{j_1,w_1}{}^{s_2,\alpha_2,w_2}}{1+e^{-2\pi i\left(\alpha_2-is_2\right)}}~~ + ~~~~(s_2\rightarrow -s_2) \, .
\eea    

Now notice that, for $w\leq-1$, the integral over $j$ can be replaced by an integral over $-\frac{k-1}{2}+is$ minus an integral over $-\frac12+is$ with $s
 \in [-\infty,0]$. For $w\geq1$,
 the original integral splits into the same two integrals, but now 
with $s \in [0,\infty ]$. Adding these terms to (\ref{ST3a})
 one ends, after some extra contour deformations in the remaining integrals,
 with 
$~~S_{j_1,w_1}^{s',\alpha',w'}$ 
and we can conclude that $({\mathcal S} T)^3_{~j,w}{}^{s',\alpha',w'}=0$.

\section{A generalized Verlinde formula}

As is well known, the Verlinde theorem allows to compute the fusion
coefficients in RCFT as:
\bea
{\cal N}_{{\mu\nu}}{}^{\rho}
=\sum_{\kappa}\frac{S_{\mu}{}^{\kappa} S_{\nu}{}^{\kappa} 
\left(S_{\rho}{}^{\kappa}\right)^{-1}}{S_{0}{}^{\kappa}}\, ,\label{verlinde}
\eea
where the index ``$0$'' refers
to the representation containing the identity field. 
In the
case of the
fractional level admissible representations of the $\widehat{sl(2)}$
 affine Lie algebra, the negative integer
fusion coefficients obtained from (\ref{verlinde})
 in \cite{mat} were interpreted as 
 a consequence
of the identification $j\rightarrow -1-j$ in \cite{awa}\footnote{
Interestingly, it was shown 
in a recent
detailed study of the $\hat{sl}(2)_{k=\frac12}$ model \cite{Ridaut},
that the origin of the negative signs
is the absence of
 spectral flow images of the admissible representations  in the analysis of
 \cite{awa}.},
where it was also
shown that fusions are not allowed by the Verlinde formula
if the fields involved are not highest- or lowest-weight.
Applications to other non RCFT were discussed in 
\cite{TJ}, where generalizations of the theorem were proposed
for 
certain representations in
the Liouville theory, the H$_3^+$ model and the SL(2,$\mathbb R)$/U(1)
 coset.

In order to explore alternative expressions 
in the AdS$_3$ model, let us 
consider the more tractable finite dimensional
 degenerate representations. From the results 
for the characters  obtained in
section 2, it is natural to
propose the following generalization of
the Verlinde formula\footnote{ A similar 
expression was 
obtained in \cite{TJ} 
 for the H$_3^+$ model applying the Cardy ansatz.}

\bea
\sum_{J_3}~~{\cal N}_{J_1J_2}{}^{J_3}~~
\chi_{J_3}(\theta,\tau,0)=\sum_{w=-\infty}^{\infty}
\int^{-\frac12}_{-\frac{k-1}{2}}dj~~
\frac{S_{J_1}{}^{j,w}S_{J_2}{}^{j,w}}{S_{0}{}^{j,w}}~~
e^{2\pi i\frac k4 \frac{\theta^2}{\tau}}\chi_j^{+,w}(\frac\theta\tau,
-\frac1\tau,0)\, ,\cr&&\label{verlindedeg}
\eea
which holds for generic $(\theta,\tau)$ far from the 
points $\theta+n\tau\in{\mathbb Z},\forall n\in{\mathbb Z}$.
In order to prove it,
 notice that, in the region of the parameters where we claim it holds, 
one can neglect the
$\epsilon's$ and contact terms on both sides of the equation
and show that the fusion 
coefficients
${\mathcal N}_{J_1J_2}{}^{J_3}$ 
coincide with those 
obtained in the H$_3^+$ model, namely
\bea
{\cal N}_{J_1J_2}{}^{J_3}=\left\{\begin{array}{lcr}
1 & & |J_1-J_2|\leq J_3\leq J_1+J_2,\cr
0 & &{\rm otherwise}\, .~~~~~~~~~~
\end{array}\right.
\label{coef}
\eea
\medskip

Let us denote the $r.h.s.$ of (\ref{verlindedeg}) 
as  $I(J_1,J_2)$ and  rewrite it as
(see (\ref{tmd}))
\bea
I(J_1,J_2)&=&\sqrt{\frac{2}{k-2}}\frac{e^{\frac{2\pi i}{k-2}
\left(\frac{k-2}{2}\right)^2\frac{\theta^2}{\tau}}}{\sqrt{i\tau}~
i\vartheta_{11}(\theta,\tau)}\int_{-\infty}^{\infty}d\lambda~~
\frac{e^{\frac{2\pi i}{k-2}\frac{\lambda^2}{\tau}}e^{2\pi i\frac{\theta}\tau
\lambda}}{e^{\pi i\sqrt{\frac{2}{k-2}}\lambda}-e^{-\pi i\sqrt{\frac{2}{k-2}}
\lambda}}
\cr\cr\cr
&&\times~\left[e^{\frac{2\pi i}{k-2}N_1\lambda}+e^{-\frac{2\pi i}{k-2}N_1
\lambda}-e^{\frac{2\pi i}{k-2}N_2\lambda}-e^{-\frac{2\pi i}{k-2}N_2\lambda}
\right],
\eea
where $N_1=2(J_1+J_2+1)$ and $N_2=2(J_1-J_2)$.
Changing $\lambda\rightarrow-\lambda$ in the second and fourth terms, we get
\bea
I(J_1,J_2)=I(N_1)-I(N_2)~,~~~~~~I(N_i)=\tilde I(N_i,\theta,\tau)+ 
\tilde I(N_i,-\theta,\tau)~, \label{sum}
\eea
with
\bea
\tilde I(N_i,\theta,\tau)=\frac{\sqrt{\frac{2}{k-2}}}{\sqrt{i\tau}~
i\vartheta_{11}(\theta,\tau)}\int_{-\infty}^{\infty}d\lambda~
\frac{e^{\frac{2\pi i}{k-2}
\frac{1}{\tau}\left(\lambda+\theta\frac{k-2}{2}\right)^2}e^{\pi 
i\sqrt{\frac{2}{k-2}}N_i\lambda}}{e^{\pi i\sqrt{\frac{2}{k-2}}
\lambda}-e^{-\pi i\sqrt{\frac{2}{k-2}}\lambda}}\, .\label{sinn}
\eea
The divergent terms in this expression cancel in the sum (\ref{sum}).

Without loss of generality, let us assume 
$J_1\geq J_2$. To perform the $\lambda$-integral in (\ref{sinn}), it is 
convenient to
split the cases with
odd and even $N_i$.
Writing
$N_i+1=2m_i$, $m_i\in\mathbb N$,  in the first case
we get
\bea
\tilde I(N_i,\theta,\tau)=\sum_{L=0}^{m_i-1}\frac{e^{\frac{-2\pi i}{k-2}
\tau L^2}e^{-2\pi i \theta L}}{i\vartheta_{11}(\theta,\tau)}-\frac{e^{\pi i
\frac{k-2}{2}\frac{\theta^2}{\tau}}}{\sqrt{i\tau}~i\vartheta_{11}
(\theta,\tau)}\int_{-\infty}^{\infty}d\lambda~~\frac{e^{\pi i\frac{\lambda^2}
{\tau}}e^{2\pi i\sqrt{\frac{k-2}{2}}\frac{\theta\lambda}{\tau}}}{1-e^{2\pi i
\sqrt{\frac{2}{k-2}}\lambda}}\, ,
\eea
where the second term diverges.
For even $N_i$, take $N_i+2=2n_i$ with $n_i\in\mathbb N$, and then
\bea
\tilde I(N_i,\theta,\tau)&=&\sum_{L=0}^{n_i-1}\frac{e^{\frac{-2\pi i}{k-2}\tau 
\left(L-\frac12\right)^2}e^{-2\pi i \theta \left(L-\frac12\right)}}
{i\vartheta_{11}(\theta,\tau)}\cr\cr
&&-~ \frac{e^{\pi i\frac{k-2}{2}\frac{\theta^2}{\tau}}}{\sqrt{i\tau}~
i\vartheta_{11}(\theta,\tau)}\int_{-\infty}^{\infty}d\lambda~~
\frac{e^{\pi i\frac{\lambda^2}{\tau}}e^{2\pi i\sqrt{\frac{k-2}{2}}
\frac{\theta\lambda}{\tau}}e^{-\pi i\sqrt{\frac{2}{k-2}}\lambda}}
{1-e^{2\pi i\sqrt{\frac{2}{k-2}}\lambda}}\, ,
\eea
where again the second term diverges.

Notice that
$N_1$ and $N_2$ are either both even or odd, and since the divergent term
is the same in 
$I(N_1)$ and $I(N_2)$, it cancels in the sum
$I(J_1,J_2)$. Thus,
putting all together we get
\bea
I(J_1,J_2)= \sum_{J_3=J_1-J_2}^{J_1+J_2}\frac{-e^{\frac{-2\pi i}{4(k-2)}
\tau(2J_3+1) ^2}2~\sin\left(\pi i \theta (2J_3+1)\right)}
{\vartheta_{11}(\theta,\tau)}=\sum_{J_3=J_1-J_2}^{J_1+J_2}\chi_{J_3}
(\theta,\tau,0)\,.
\eea
where we have defined $J_3=L-\frac12$ for odd $N_1$ and $N_2$ and 
$J_3=L-1$ for  even $N_1$ and $N_2$.

From a similar analysis of the case $J_2>J_1$, we obtain
(\ref{verlindedeg}) and (\ref{coef}). 

 In conclusion, consistently 
with the assumption that correlation functions of fields in 
degenerate representations in the H$_3^+$ and AdS$_3$ models
 are 
related by analytic continuation,
the generalized Verlinde formula (\ref{verlindedeg})
reproduces the  fusion rules of degenerate representations
previously obtained in the Euclidean model.
However,  even if it is not expected to reproduce
the fusion rules of  continuous representations \cite{awa},
applying it for discrete representations
 also fails.

\section{One-point functions}

In this appendix we summarize
 the results for one-point functions in
 maximally symmetric D-branes, obtained by applying the method of 
\cite{israel}. The solution for 
 one-point functions in H$_2$ D-branes found in $loc. cit.$ 
holds for integer level $k$. 
Here we work with an alternative expression, equivalent to the one obtained 
in \cite{israel}, but with a different extension for generic $k\in{\mathbb R}$.

The method 
rests on the observation that, after doing a T duality in the timelike 
direction, the 
$N$-th cover of SL(2,${\mathbb R})$, $i.e.$
SL(2,${\mathbb R})^N_k$, is given by the orbifold
\bea
\frac{SL(2,{\mathbb R})_k/U(1)\times U(1)_{-k}}{{\mathbb Z}_{Nk}}\, .
\eea

Because now the timelike direction is a free compact boson, 
the analytic continuation to  Euclidean space 
is simply obtained by replacing $U(1)_{-k} \rightarrow U(1)_{R^2k}$. 
Thus, one can construct arbitrary correlation functions in 
AdS$_3$ from those in the cigar and  the free compact boson theories,
after taking 
the limits 
$N\rightarrow\infty$, $R^2\rightarrow-1$. The effect of the orbifold is to
 produce new (twisted) sectors. 
 These can be read in the following modification of the left and right 
momentum modes in the coset 
and the free boson models, respectively,  
\bea
\frac{\left(n+k\omega,n-k\omega\right)}{\sqrt{2k}}&\longrightarrow& 
\frac{\left(n+k\omega-\frac{\gamma}{N},n-k\omega+\frac{\gamma}{N} \right)}
{\sqrt{2k}},~~ \gamma\in{\mathbb Z}_{kN}\, ,\cr
\frac{\left(\tilde n+R^2k\tilde\omega,\tilde n-R^2k\tilde \omega\right)}
{R\sqrt{2k}}&\longrightarrow& \frac{\left(n+kNp+R^2k\tilde 
\omega+\frac{R^2\gamma}{N},n+kNp-R^2k\tilde \omega-\frac{R^2\gamma}{N} 
\right)}{R\sqrt{2k}}
\, ,\cr&&
\eea
with $p\in{\mathbb Z}$ and $\omega, \tilde\omega$ being the 
winding numbers in the cigar and U(1) respectively.
In the $N$-th cover, 
$k$ has to be an integer, but in the universal 
covering,
the theory can be defined for arbitrary real level $k>2$ \cite{israel}.

The vertex operators for the orbifold theory are the product of the 
vertices in each space, namely
\bea
V^j_{n\omega\gamma p\tilde\omega}(z,\bar z)=\Phi^{sl(2)/u(1)}_{j,n,
\omega-\frac{\gamma}{kN}}
(z,\bar z)~~\Phi^{u(1)}_{n+kNp,\tilde\omega+\frac{\gamma}{kN}}(z,\bar z)\, .
\label{ver}
\eea
In the universal covering, 
the discrete momentum $\frac{\gamma}{kN}$ becomes a continuous parameter 
$\lambda\in [0,1)$,
the $J_0^3,\bar J_0^3$ quantum numbers read
\bea
M=-\frac n2+\frac k2(\tilde\omega+\lambda),~~~~ \bar M=\frac n2 +\frac 
k2(\tilde\omega+\lambda),
\eea 
and the winding number is given by 
\bea
w=\omega+\tilde\omega.
\eea

\subsection{One-point functions for point-like instanton branes}

To obtain the one-point functions for the point-like branes, 
we simply take the ${\mathbb Z}_{kN}$ orbifold 
action on the product of the one-point functions  associated 
to D0  
branes in the cigar
\cite{RS} 
and to Neumann boundary conditions in the U(1) theories, 
respectively
\bea
&&\left\langle \Phi_{j,n,\omega}^{sl(2)/u(1)}(z,\bar z)\right\rangle_{\bf s}
^{D0} =
\frac{\delta_{n,0}~(-)^{r\omega}}
{\left|z-\bar z\right|^{h^j_{nr}+\bar h^j_{nr}}} 
\frac{\Gamma\left(-j+\frac k2 \omega\right)
\Gamma\left(-j-\frac k2 \omega\right)}{\Gamma\left(-2j-1\right)}\cr\cr
&&  \ \ \ \ \ \ \ \ \ \ \ 
\times~\left(\frac{k}{k-2}\right)^{\frac14}\left(\frac{ \sin[\pi b^2]}
{4\pi}\right)^{\frac12}\frac{\sin[{\bf s}(2j+1)]}{\sin[\pi b^2(2j+1)]}
\frac{\Gamma\left(1+b^2\right)~\nu^{1+j}}{\Gamma\left(1-b^2(2j+1)\right)}
\, ,
\label{D0}
\eea 
and
$\ \ \ \ \ \ \ \ \ \ \ \ \ \ \ \ \ \ 
\displaystyle
\left\langle \Phi^{u(1)}_{\tilde n,\tilde \omega}(z,\bar z)\right
\rangle_{x_0}^{\cal N} ~= ~
\frac{\delta_{\tilde n,0}e^{i\tilde \omega x_0}
\left(\sqrt{k/2}R\right)^{\frac12}}{\left|z-\bar z\right|^{\frac k2 
\tilde\omega^2}}\, .
$

\medskip

Here ${\bf s}
=\pi rb^2,~r\in{\mathbb N}$, $b^2=\frac{1}{k-2}$, $\nu=\pi\frac{\Gamma\left(1-\frac{1}{k-2}\right)}{\Gamma\left(1+\frac{1}{k-2}\right)}$, ${\mathcal N}$ refers to 
Neumann boundary 
conditions\footnote{Recall that we considered 
Dirichlet gluing conditions when 
constructing the coherent states. Here
we take Neumann boundary conditions because 
this is the T dual version in the time direction.} 
and $x_0$ is 
the 
position of the D0 brane in the timelike direction. In the single covering
of 
SL(2,${\mathbb R})$, the only possibilities are 
$x_0=0$ and $\pi$, which represent the 
center of the group ${\mathbb Z}_2$ (see \cite{Stanciu}). But in the universal 
covering, one can take $x_0=q\pi$ with $q\in {\mathbb Z}$ 
(see section \ref{conjclass}).   

To compare these one-point functions with those obtained 
in section \ref{Dbranes}, it is convenient to 
consider the conventions used in \cite{wc}~\footnote{Notice that here we take a different normalization in order to explicitly realize the relation between the spectral flow image of highest and lowest weight representations.}. 
There, the fields $\Phi_{m,\bar m}^{j,w}$ represent the 
spectral flow images of the primary fields
$\Phi_{m,\bar m}^{j,0}$, $i.e$ 
they are  in correspondence with  highest or lowest weight states
depending if $w<0$ or $w>0$, and have $J_0^3,~\bar J_0^3$ 
eigenvalues $M=m+\frac k2 w,~\bar M=\bar m+\frac k2 w$. 
They are related to the vertex operators (\ref{ver}) as
\bea
\Phi_{m,\bar m}^{j,w}(z,\bar z)=(-)^w \sqrt{B(j)}~~V^{-1-j}_{n\omega\gamma p\tilde \omega}
(z,\bar z)\,,~~~~~~B(j)=\frac{k-2}{\pi}~\frac{\Gamma\left(1+\frac{1+2j}{k-2}\right)}{\Gamma\left(-\frac{1+2j}{k-2}\right)}~\nu^{\frac12+j}, ~~
\eea 

When looking for $w=0$ solutions, $i.e.$ $\omega=-\tilde\omega$, one expects 
to reproduce the  one-point functions of point-like D-branes
in the H$_3^+$ model, which forces $x_0=r\pi$. So,
\bea
\left\langle \Phi_{m,\bar m}^{j,w}(z,\bar z)\right\rangle_{\bf s} &=& 
\frac{\delta_{m,\bar m}}{\left|z-\bar z\right|^{\Delta_j+\bar\Delta_j}}
~\frac{\Gamma\left(1+j-m\right) \Gamma\left(1+j+m\right)}{\Gamma
\left(2j+1\right)}\cr\cr
&&\times ~\frac{i\sqrt k~(-)^{w+1}}{2^{\frac54}}~ \frac{\sin[{\bf s}\left((2j+1)-w(k-2)\right)]}{\sqrt{\sin[\frac{\pi}{k-2}\left(2j+1\right)]}}\label{final}
~,
\eea
with the parameter ${\bf s}$ labeling the positions of the instanton 
solutions. 

Comparing the OPE
\bea
J^3(\zeta)~\Phi^{j,w}_{m,\bar m}\left(\xi,\bar \xi\right)&=& 
\frac{m+\frac k2 w}{\zeta-\xi}~\Phi^{j,w}_{m,\bar m}
\left(\xi,\bar \xi\right)+\dots\cr\cr
J^{\pm}(\zeta)~\Phi^{j,w}_{m,\bar m}\left(\xi,\bar \xi\right)&=& 
\frac{m\mp j}{(\zeta-\xi)^{1\pm w}}~\Phi^{j,w}_{m,\bar m}
\left(\xi,\bar \xi\right)+\dots
\eea
and the 
antiholomorphic ones with those of the fields 
$\Phi^{(P)}$ of section \ref{Dbranes}, namely  (\ref{ope(p)}), 
we obtain the following relation, 
valid for $m=\bar m\in-j+{\mathbb Z}_{\geq0}$,
\bea
\Phi^{j,w}_{m,\bar m}(\xi,\bar \xi)= \Omega~ (-)^{j+m}
\frac{\Gamma\left(1+j-m\right)\Gamma\left(1+j+m
\right)}{\Gamma\left(1+2j\right)}~\Phi^{(P)}
\left(\left|j,m,\bar m,w\right\rangle;\xi,\bar \xi\right)\, ,\label{norm}
\eea 
where $\Omega$ is the normalization of $\Phi^{j,w}_{-j,-j}$. We find perfect 
agreement between the expressions 
(\ref{resul}) and (\ref{final}) for one-point functions, as long as 
$~~\Omega=-\sqrt{\frac{-ik(k-2)}{16}}$ .

\subsection{One point-functions for H$_2$, dS$_2$ and light-cone branes} 
 
All of the H$_2$, dS$_2$ and light-cone  branes can be constructed from a 
D2-brane in the 
cigar and taking Neumann boundary conditions in the U(1). 
They are simply related to each other
by analytic continuation of a parameter 
labeling the scale of the branes. 
Here, we  discuss in 
detail the case of the one-point functions of fields in 
discrete representations on H$_2$ branes 
and show that 
the {\it Cardy structure} is realized in this case. These one-point functions 
correspond to H$_2$ branes at $X^3=cons$ rather than $X^0=cons$, so we have
to translate these solutions before comparing with the results of section
\ref{Dbranes}.

The one point-functions for the D2-branes in the cigar are given by \cite{RS} 
\bea
\left\langle \Phi_{j,n,\omega}^{sl(2)/u(1)}(z,\bar z)
\right\rangle_{\tilde\sigma}^{D2} &=& 
\frac{\frac12 \delta_{n,0} (-)^\omega e^{-i\tilde\sigma\omega(k-2)}
\left(\frac{k-2}{k}\right)^{\frac14}}{\left|z-\bar z\right|^{h^j_{nr}+
\bar h^j_{nr}}} 
\Gamma\left(1+2j\right)\Gamma\left(1+\frac{1+2j}{k-2}\right)
\nu^{\frac12+j}\cr\cr\cr
&&\times ~\left(\frac{\Gamma\left(-j+\frac k2 \omega\right)}{\Gamma\left(1+j+
\frac k2\omega\right)}e^{i\tilde\sigma(1+2j)}+ 
\frac{\Gamma\left(-j-\frac k2 \omega\right)}
{\Gamma\left(1+j-\frac k2\omega\right)}e^{-i\tilde\sigma(1+2j)}\right)\, .
\nonumber
\eea 
Notice that this differs from the result
 in \cite{RS} by the $\omega$ 
dependent phase 
$(-)^\omega~e^{-i\tilde\sigma\omega(k-2)}$.\footnote{ This phase 
that we added by hand is required by the spectral flow symmetry, 
when used to construct the one-point functions for H$_2$ branes, 
which demands
$\left\langle 
\Phi_{j,j}^{j,w}\right\rangle^{H_2}=\left\langle 
\Phi_{\frac k2+j,\frac k2+j}^{-\frac k2-j,w-1}\right\rangle^{H_2}$,
 in our conventions.
 The one-point function for D2 branes 
was constructed in \cite{RS}
beginning from the parent 
H$_3^+$ model  and was found to have some sign 
problems. We claim this phase cannot be deduced  from the H$_3^+$ model 
because of the absence of spectral flowed states. It would be interesting to 
investigate the implications of this modification in the sign. 
Unfortunately, this information cannot be obtained from the $w$
 independent
 semiclassical limit of the one-point functions.}
The position of the D-brane over the $U(1)$ is again fixed 
by the one-point function of 
the H$_3^+$ model. We find
\bea
\left\langle \Phi_{m,\bar m}^{j,w}(z,\bar z)\right\rangle_{\tilde\sigma}^{H_2,X^3} &=& 
\frac{\delta_{m,\bar m}}{\left|z-\bar z\right|^{\Delta_j+\bar 
\Delta_j}}\frac{-1}{2^{\frac54}~\sqrt{i} (k-2)^{\frac14}}\frac{\pi ~e^{-i\tilde\sigma w(k-2)}}{\sqrt{\sin\left[\frac{\pi}{k-2}(2j+1)\right]}}\cr\cr\cr
&\times & \left(\frac{\Gamma\left(1+j-m\right)}{\Gamma\left(-j-m\right)}
e^{-i\tilde\sigma(1+2j)}+ 
\frac{\Gamma\left(1+j+m\right)}{\Gamma\left(-j+m\right)}
e^{i\tilde\sigma(1+2j)}\right)\, .~~~~
\label{1ptH2}
\eea

For fields in discrete representations with 
$m=-j+{\mathbb Z}_{\geq0}$ and $j\notin{\mathbb Z}$, only one factor survives in the  last line. 
Here $\tilde\sigma$ is a real parameter, determining the embedding   of
 the brane in AdS$_3$ as $X^3=\cosh\rho \sin\tau=\sin\tilde\sigma$.
So, in order to compare with the solutions discussed in section \ref{Dbranes}, 
the identification $\tilde\sigma=\sigma+
\frac\pi2$ and the global shift in the time-like coordinate on the cylinder, 
namely  $t\rightarrow t+\frac\pi2$,
must be perfomed. The latter simply adds a phase 
$e^{i\frac\pi2(M+\bar M)}$ (in fact, $J_0^3+\bar J_0^3$ gives the energy in 
AdS$_3$ and so this combination is the generator of $t$ translations). 

From the analysis of conjugacy classes, it is natural to relabel 
$\sigma=\frac{\pi}{k-2}(2j'+1)-w'\pi$, with $j'\in
(-\frac{k-1}{2},-\frac12)$, $w'\in{\mathbb Z}$
\footnote{The one-point functions for dS$_2$ branes are 
given by (\ref{1ptH2}) with $j'\in\left\{-\frac12+i{\mathbb R^+}\right\}$ and 
for light-cone branes, they are given by $\sigma=n\pi,\,n\in{\mathbb Z}$.}, and 
\bea
\left\langle \Phi_{m,\bar m}^{j,w}(z,\bar z)\right
\rangle_{\sigma(j',w')}^{H_2,X^0} &=&
\frac{\delta_{m,\bar m}}{\left|z-\bar z\right|^{\Delta_j+\bar 
\Delta_j}} ~ \frac{\Gamma\left(1+j+m\right)\Gamma\left(1+j-m\right)}
{\Gamma\left(1+2j\right)}
\cr\cr
&&\times ~ \frac{-\pi \sqrt{-i}}{2^{\frac54}(k-2)^{\frac14}}\frac{(-)^w e^{\frac{4\pi i}{k-2}(j'+\frac12-w'\frac{k-2}{2})(j+\frac12-w
\frac{k-2}{2})}}{\sqrt{\sin\left[\frac{\pi}{k-2}(2j+1)\right]}}\,.\label{H2dis}
\eea 

\subsection{One-point functions for AdS$_2$ branes}

For completeness, we display here
the one-point functions for AdS$_2$ branes obtained
in \cite{israel}, in our conventions.
These
 are constructed by gluing two one-point functions: one for a  D1-brane
in the coset model
and another one with Dirichlet boundary conditions in the U(1) model. 
The result is
\bea
\left\langle \Phi^{j,w}_{m,\bar m}(z,\bar z)\right\rangle^{AdS_2}_r &=& 
\frac{\delta_{w,0}\delta_{m,-\bar m} e^{-i\frac\pi4}e^{in(\theta_0+x_0)}
\left(\frac{k-2}{2}\right)^{\frac14}}{\left|z-\bar z\right|^{\Delta_j+\bar
\Delta_j}}
\frac{\Gamma\left(-1-2j\right)}{\Gamma\left(-j-m\right)\Gamma
\left(-j+m\right)}\cr\cr\cr
&&\times ~\cos\left(ir(j+\frac12)+m\pi\right)
 \Gamma\left(1-\frac{1+2j}{k-2}\right)\nu^{-\frac12-j}\, ,\label{gam}
\eea
where $\theta_0$ is related to the angles (in cylindrical coordinates) 
to which the branes asymptote when they get close to the boundary of AdS$_3$,
 $x_0$ is the location of the brane and $r$ determines their scale. 
From the geometrical point of view, $r$ seems to be an arbitrary real number, 
but as  shown in \cite{BP}, it becomes quantized at the semiclassical level.
 
Let us end this appendix by noticing the perfect agreement 
with the analysis of the
coherent states presented in section 4. Due to the
 Gamma-functions in the
denominator of (\ref{gam}),
only states in the
continuous representations couple to the AdS$_2$ branes
and, due to the delta-functions, 
only those with $w=0$ and  
$m=-\bar m$ have non vanishing expectation values.


\bigskip


\begin{thebibliography}{99}
\bibitem{mo1} J. Maldacena and H. Ooguri, {\it
Strings in AdS$_3$ and
  the $SL(2,\mathbb R)$ WZW Model: Part 1: The Spectrum},
  J. Math. Phys. {\bf 42}, 2929 (2001); [arXiv:0001053 [hep-th]].
\bibitem{mo2} J. Maldacena, H. Ooguri and J. Son, 
{\it Strings in $AdS_3$ and the SL(2,R) WZW model.
Part 2: Euclidean Black Hole}, J. Math. Phys. {\bf 42}, 2961 (2001); 
[arXiv:0005183 [hep-th]].
\bibitem{mo3}
J. Maldacena and H. Ooguri, {\it Strings in $AdS_3$ and the
$SL(2, \mathbb {R})$ WZW Model. Part 3: Correlation Functions},
Phys. Rev. {\bf D65}, 106006 (2002); [arXiv:011180 [hep-th]].
\bibitem{wc} W. Baron and C. N\'u\~nez, 
{\it Fusion rules and four-point functions in the AdS$_3$ WZNW model},
Phys. Rev. {\bf D79}, 086004 (2009); [arXiv:0810.2768 [hep-th]].
\bibitem{tesch1} J. Teschner, {\it On structure constants and fusion rules
in the $SL(2, \mathbb {C})/SU(2)$ WZW model}, Nucl. Phys. {\bf B546}, 
 390 (1999);
 [arXiv:9712256 [hep-th]].

\bibitem{tesch3} J. Teschner,
{\it Operator product expansion and factorization
in the H$^+_3$ WZW model}, Nucl. Phys. {\bf B571}, 
555 (2000);
[arXiv:9906215 [hep-th]].

\bibitem{rib} S. Ribault, {\it Minisuperspace limit of the AdS$_3$ WZNW model},
JHEP {\bf 1004}, 096 (2010); [arXiv:0912.4481 [hep-th]].


\bibitem{verlinde} E. Verlinde, {\it Fusion rules and modular transformations
in conformal field theory}, Nucl. Phys. {\bf B300}, 360 (1988).
\bibitem{TJ} C. Jego and J. Troost, {\it Notes on the Verlinde formula in 
non rational conformal field theories},  Phys. Rev. {\bf D74}, 106002 
(2006); [arXiv: 0601085 [hep-th]]. 

\bibitem{stt} A. M. Semikhatov, A. Taormina and I. Yu. Tipunin, {\it
 Higher-level appell functions, modular transformations and characters}, 
arXiv:math/0311314.
\bibitem{est} T. Eguchi, Y. Sugawara and A. Taormina, {\it Liouville field,
modular forms and elliptic genera}, JHEP {\bf 0703}, 119 (2007); 
[arXiv:0611338 [hep-th]].
\bibitem{t} A. Taormina, {\it Liouville theory and elliptic genera}, 
Prog. Theor. Phys. Suppl. {\bf 177}, 203 (2009);
[arXiv:0808.2376 [hep-th]].
\bibitem{hhrs} M. Henningson, S. Hwang, P. Roberts and B. Sundborg, {\it
Modular invariance of SU(1,1) strings}, Phys. Lett. {\bf B267}, 350 (1991).

\bibitem{HS} Y. Hikida, Y. Sugawara, {\it Boundary states of D branes in 
AdS(3) based on discrete series},  Prog. Theor. Phys. {\bf 107}, 1245 
(2002); [arXiv: 0107189 [hep-th]]. 

\bibitem{gawe}K. Gawedski, {\it Noncompact WZW conformal field theories},
Proceedings of the NATO Advanced Study Institute, {\it New Symmetry
Principles in Quantum Field Theory}, Cargese, 1991, p. 247, eds. J. Frolich, G.
´t Hooft, A. Jaffe, G. Mack, P.K. Mitter and R. Stora, Plenum Press 1992;
[arXiv:9110076 [hep-th]].
\bibitem{israel} D. Israel, {\it D-branes in Lorentzian $AdS_3$}, JHEP 
{\bf 0506}, 008 (2005); [arXiv: 0502159 [hep-th]].


\bibitem{kounnas} D. Israel, C. Kounnas and P. Petropoulos, 
{\it Superstrings on NS5 backgrounds, deformed AdS$_3$ and holography},
JHEP {\bf 0310}, 028 (2003); [arXiv:0306053 [hep-th]].

\bibitem{hpt}  A. Hanany, N. Prezas and J. Troost, {\it The partition 
function of the two-dimensional black hole conformal field theory}, JHEP
{\bf 0204},  014  (2002)
[arXiv:0202129 [hep-th]]




\bibitem{GKS} A. Giveon, D. Kutasov, A Shwimmer, {\it Comments on D-branes 
in $AdS_3$}, Nucl. Phys. {\bf B615}, 133 (2001); [arXiv: 0106005 [hep-th]].
\bibitem{PST} B. Ponsot, V. Schomerus, J. Teschner, {\it Branes in 
the Euclidean $AdS_3$}, JHEP 0202, 016 (2002); [arXiv:0112198 [hep-th]]. 












\bibitem{Stanciu} S. Stanciu, {\it D-branes in an $AdS_3$ background}, 
JHEP {\bf9909}, 028 (1999); [arXiv: 9901122 [hep-th]].

\bibitem{FS} J. M. Figueroa-O'Farrill and S. Stanciu,
 {\it D-branes in $AdS_3\times S_3\times S_3\times S_1$ background}, 
JHEP {\bf 0004}, 005 (2000); [arXiv:0001199 [hep-th]].

\bibitem{BP} C. Bachas, M. Petropoulos, {\it Anti-de Sitter D-branes}, JHEP 
{\bf 0102}, 025 (2001); [arXiv: 0012234 [hep-th]]. 

\bibitem{PR} P.M. Petropoulos, S. Ribault, {\it Some comments on Anti-de 
Sitter D-branes}, JHEP {\bf 0107}, 036 (2001); [arXiv: 0105252 [hep-th]].


\bibitem{LOPT} P. Lee, H. Ooguri, J. Park, J. Tannenhauser, 
{\it Open strings on $AdS_2$ branes}, Nucl. Phys. {\bf B610}, 3 (2001); 
[arXiv: 0106129 [hep-th]].

\bibitem{LOP} P. Lee, H. Ooguri, J. Park, {\it Boundary states for $AdS_2$ 
branes in $AdS_3$}, Nucl.Phys. {\bf B632}, 283 (2002); 
[arXiv:0112188 [hep-th]].



\bibitem{PS} A. Parnachev, D. Sahakyan, {\it Some remarks on D-branes in 
$AdS_3$}, 
JHEP {\bf 0110}, 022 (2001); [arXiv: 0109150 [hep-th]].

\bibitem{RR} A. Rajaraman and M. Rozali, {\it Boundary States for D-branes in 
$AdS_3$}, Phys. Rev. {\bf D66}, 026006 (2002); [arXiv:0108001 [hep-th]].

\bibitem{D}  C. Deliduman, {\it $AdS_2$ D-branes in Lorentzian $AdS_3$}, 
Phys. Rev. {\bf D68}, 066006 (2003); [arXiv: 0211288 [hep-th]].

\bibitem{H} W. H. Huang, {\it Anti-de Sitter D-branes in Curved 
Backgrounds}, JHEP {\bf 0507}, 031 (2005); [arXiv: 0504013 [hep-th]].
 
\bibitem{RS} S. Ribault, V. Schomerus, 
{\it Branes in the 2-D Euclidean Black hole}, 
JHEP {\bf 0402}, 019 (2004); [arXiv: 0310024 [hep-th]]. 
 
\bibitem{AS} A. Y. Alekseev, V. Schomerus, {\it D-branes in the WZW model}, 
Phys. Rev. {\bf D60}, 061901 (1999); [arXiv: 9812193 [hep-th]]. 

\bibitem{dif} P. Di Francesco, P. Mathieu and D. S\'en\'echal, {\it Conformal
Field Theory},  Springer-Verlag, New York, 1997.


\bibitem{KO}M. Kato, T. Okada, {\it D-branes on group manifolds}, 
Nucl. Phys. {\bf B499}, 583 (1997); [arXiv:9612148 [hep-th]]. 

\bibitem{Ishibashi} N. Ishibashi, {\it The Boundary and Crosscap States in 
Conformal Field Theories}, Mod. Phys. Lett. {\bf A4}, 251 (1989). 



\bibitem{schomerus} Volker Schomerus, {\it Lectures on branes in curved 
backgrounds}, Class. Quant. Grav. {\bf 19}, 5781 (2002);
 [arXiv:0209241 [hep-th]].



\bibitem{FZZ} V. Fateev, A. Zamolodchikov, Al. Zamolodchikov, 
{\it Boundary Liouville field theory. 1. Boundary state and boundary 
two point function}, 
arXiv: 0001012 [hep-th].


\bibitem{moore} G. Moore, {\it Finite in all directions}, 
arXiv:hep-th/9305139.
\bibitem{seiberg} H. Liu, G. Moore and N. Seiberg, {\it Strings in a time
dependent orbifold}, JHEP {\bf 0206}, 045 (2002); [arXiv:0204168 [hep-th]].
\bibitem{kutasov} B. Craps, D. Kutasov and G. Rajesh, {\it String
propagation in the presence of cosmological singularities}, JHEP {\bf 0206},
053 (2002);
[arXiv:0205101 [hep-th]].
\bibitem{russo} G. Papadopoulos, J. Russo and A. Tseytlin, {\it Solvable models
of strings in a time-dependent plane-wave background}, 
Class. Quant. Grav. {\bf 20}, 969-1016 (2003); [arXiv:0211289 [hep-th]].


\bibitem{mat} P. Mathieu and M. Walton, Prog. Theor. Phys. Suppl.
{\bf 102},
229  (1990).

\bibitem{awa} H. Awata and Y. Yamada, {\it Fusion rules for the fractional
level ${\widehat sl(2)}$ algebra}, Mod. Phys. Lett. {\bf A7}, 1185 (1992).

\bibitem{gk} A. Giveon and D. Kutasov, {\it Notes on $AdS_3$},
  Nucl. Phys. {\bf B621},
303  (2002) ; [arXiv:0106004 [hep-th]]. 



\bibitem{hs} K. Hosomichi and Y. Satoh, {\it Operator product
  expansion in SL(2) conformal field theory}, Mod. Phys. Lett. {\bf
  A17}, 683 (2002); [arXiv:0105283 [hep-th]].

\bibitem{Ridaut} D. Ridout, {\it $\hat{sl}(2)_{-1/2}$: A Case Study}, Nucl. 
Phys. {\bf B814}, 485 (2009); [arXiv:0810.3532 [hep-th]].

 


\end{thebibliography}
\end{document}